\documentclass{aastex63}

\usepackage{enumerate}

\received{March 31, 2022}
\revised{May 23, 2022}
\accepted{May 23, 2022}
\shorttitle{Modeling \textit{PSP} solar encounters 1 and 2 with ADAPT-WSA}
\shortauthors{Wallace et al.}
\graphicspath{{./}{figures/}}

\begin{document}


\title{New insights into the first two PSP solar encounters enabled by \\modeling analysis with ADAPT-WSA}

\author[0000-0002-1091-4688]{Samantha Wallace}
\affiliation{NASA Goddard Space Flight Center \\
8800 Greenbelt Rd \\
Greenbelt, MD 20771, USA}

\author[0000-0001-9498-460X]{Shaela I. Jones}
\affiliation{NASA Goddard Space Flight Center \\
8800 Greenbelt Rd \\
Greenbelt, MD 20771, USA}

\author[0000-0001-9326-3448]{C. Nick Arge}
\affiliation{NASA Goddard Space Flight Center \\
8800 Greenbelt Rd \\
Greenbelt, MD 20771, USA}

\author[0000-0003-1692-1704]{Nicholeen Viall}
\affiliation{NASA Goddard Space Flight Center \\
8800 Greenbelt Rd \\
Greenbelt, MD 20771, USA}

\author[0000-0002-6038-6369]{Carl J. Henney}
\affiliation{Air Force Research Laboratory \\
Space Vehicles Directorate \\
Kirtland AFB, NM 87117, USA}

\begin{abstract}


\textit{Parker Solar Probe}'s (\textit{PSP})'s unique orbital path allows us to observe the solar wind closer to the Sun than ever before. Essential to advancing our knowledge of solar wind and energetic particle formation is identifying the sources of \textit{PSP} observations.  We report on results for the first two \textit{PSP} solar encounters derived using the Wang-Sheeley-Arge (WSA) model driven by Air Force Data Assimilative Photospheric Flux Transport (ADAPT) model maps. We derive the coronal magnetic field and the 1 \(R_\odot\) source regions of the \textit{PSP}-observed solar wind.  We validate our results with the solar wind speed and magnetic polarity observed at \textit{PSP}. When modeling results are very reliable, we derive time series of model-derived spacecraft separation from the heliospheric current sheet, magnetic expansion factor, coronal hole boundary distance, and photospheric field strength along the field lines estimated to be connected to the spacecraft. We present new results for Encounter 1, which show time evolution of the far-side mid-latitude coronal hole that \textit{PSP} co-rotates with.  We discuss how this evolution coincides with solar wind speed, density, and temperature observed at the spacecraft. During Encounter 2, a new active region emerges on the far-side, making it difficult to model. We show that ADAPT-WSA output agrees well with \textit{PSP} observations once this active region rotates onto the near-side, allowing us to reliably estimate the solar wind sources retrospectively for most of the encounter. We close with ways in which coronal modeling enables scientific interpretation of these encounters that would otherwise not have been possible.

\end{abstract}


\section{Introduction} \label{sec:1_Intro}

The \textit{Parker Solar Probe} (\textit{PSP}: \citealp{Fox2016}) is a historic heliospheric mission launched in 2018 that provides unprecedented observations of the Sun and solar wind. The mission is executed in a series of close encounters with the Sun, in which seven Venus gravity assists are performed to gradually reduce the spacecraft's perihelion from $\sim$35.6 \(R_\odot\) to  $\sim$9.89 \(R_\odot\) by 2025. This unique orbit allows the spacecraft to simultaneously observe the corona remotely while probing the solar wind in situ closer to the Sun than ever before. At these radial distances \textit{PSP} measures pristine solar plasma outflow before it has had time to evolve significantly.  Each encounter also includes a unique period in which \textit{PSP} co-rotates with the Sun allowing the spacecraft to observe the solar wind emerging from the same source region over several days.  \textit{PSP} has fortuitously come online alongside other historic observatories, such as \textit{Solar Orbiter} (\textit{SolO}: \citealp{SolO-Muller2020}) and the \textit{Daniel K. Inouye Solar Telescope} (\textit{DKIST}: \citealp{DKIST-Rimmele2020}) which collectively will revolutionize our understanding of the Sun-Heliosphere system.

\textit{PSP} has three primary science objectives, none of which can be fully realized without accurate knowledge of the precise source regions of the observed solar wind, and the magnetic connectivity between the spacecraft and those sources:

\begin{enumerate}

\item Determine the mechanism behind coronal heating and the acceleration of the solar wind.

\item  Characterize the magnetic structure and plasma dynamics of the solar wind at its sources.

\item  Explore the mechanisms that accelerate and transport energetic particles from the Sun.

\end{enumerate}

Historically, relating the in situ observed solar wind back to its source has fundamentally shaped our understanding of how the solar wind is formed. By comparing in situ solar wind observations with remote observations of the corona, we know that faster (v $>$ 500 km/s) wind emerges from cold, more tenuous coronal holes (CHs) while slower (v $<$ 500 km/s) wind emerges from the ``streamer belt'' \citep{McComas2003} where coronal plasma is hotter and denser.  This region can span nearly all heliolatitudes during periods of high solar activity \citep{McComas2007} and consists of dipolar helmet streamers and unipolar ``pseudo''-streamers. The defining feature of the streamer belt is the heliospheric current sheet (HCS) formed by large helmet streamers, where the global coronal field changes sign overall.  Similarly, we relate in situ observations of elemental abundances with that observed in the corona remotely to distinguish between solar wind that originates from magnetically ``open'' flux tubes (\textit{i.e.} deep inside a coronal hole), or ``closed'' loops that release plasma at the magnetic open-closed boundary of a coronal hole \citep{Laming2015}. 


We are now in an era in which we resolve the solar surface (or photosphere) down to just 30 km (\textit{i.e.} roughly 3 orders of magnitude smaller than a supergranule), and the inner heliosphere to kinetic scales in situ. However, even in this era of unprecedented and detailed observations, it is not possible to link remote and in situ data to resolve solar wind sources beyond broad regions of the solar atmosphere (\textit{i.e.} streamer belt vs. coronal hole, open or closed field) without the use of a predictive coronal model.  Having the capability to determine the specific source locations at 1 \(R_\odot\) of the observed solar wind (\textit{i.e.} which coronal hole or coronal hole boundary) and the associated local magnetic field  (\textit{i.e.} active region or quiet Sun) is critical to the \textit{PSP} mission for several reasons. First, coronal heating and solar wind acceleration are empirically known to be related to the local magnetic field at the solar source region, each producing a different type of solar wind.  For example, the magnetic open-closed boundaries of coronal holes are sources of slower solar wind likely released from interchange reconnection \citep{Antiochos2011}. However, these boundaries are comprised both active region and quiet Sun plasma, of which active region plasma is typically hotter and more dense.  On the contrary, solar wind originating from deep within a coronal hole may be accelerated mostly by wave-turbulent driven processes \citep{Cranmer2007}.  Any reconnection that contributes to heating, release, and acceleration deep within coronal holes would be due to either open-open reconnection \citep{Tenerani2016} or interchange reconnection involving much shorter loops \citep{Fisk1999,Fisk2003}.

Second, using coronal models to identify where the observed solar wind came from on the Sun allows us to separate out and characterize solar wind originating from specific source regions (\textit{e.g.} active regions vs. quiet Sun magnetic field, coronal hole boundaries vs. deep inside CHs).  This enables us to interpret in situ observations based on the magnetic structure and plasma dynamics at the solar source region, which has implications for the physical mechanisms (\textit{e.g.} waves, reconnection) driving the formation of the solar wind from each source, thus fulfilling the second objective. For example, the streamer belt is comprised of a mix of active region and quiet Sun magnetic fields \citep{ViallBorovsky2020}, which could play an equal or greater role than the streamer type (\textit{i.e.} pseudostreamer vs helmet streamer) in how the solar wind originating from this region is heated, released, and accelerated. However, it is not always possible to differentiate between solar wind originating from these sources in situ. In situ data only reveal periods when a spacecraft crosses the HCS (via polarity reversals or electron strahl data), and offer little information regarding a spacecraft's position relative to the HCS before or after observed crossings. This is especially problematic during solar minimum when the HCS is nearly parallel to the ecliptic plane, and a spacecraft can observe HCS wind for hours to days without crossing this boundary. Knowing if the solar wind originated either near to, or far from the HCS is critical to test solar wind formation theories \citep{Antiochos2011}. This is because it is likely that HCS-related wind and the dynamics driving its formation could be vastly unique from wind that originates far from the HCS. Similarly, pseudostreamers are virtually unidentifiable in situ because they do not form a current sheet like their dipolar counterparts, and thus do not have an associated polarity inversion detectable in situ. Although both streamer types are prominent features of the coronal streamer belt and sources of the slower solar wind, it is unclear if their differing magnetic topologies result in the formation of different types of solar wind, each with their own unique physical properties and observables. Third, since solar energetic particles (SEPs) travel along magnetic field lines, knowledge of the magnetic connectivity between observed SEPs in situ and the precise field line footpoints at their source is essential to accomplish the third objective of the \textit{PSP}  mission.  Identifying the source region of such events provides key insight to the initial conditions driving their acceleration and transport.

Coronal and solar wind models have been heavily used to support the \textit{PSP} mission (e.g. \citealp{Szabo2020,Badman2020,Panasenco2020,Korreck2020,Nieves-Chinchilla2020}). The defining key to these models is that they derive the coronal magnetic field from ground and space-based remote observations of the Sun's surface magnetic field.  They also derive the sources of the in situ observed solar wind.  These models are used in two ways, each of which support the \textit{PSP} mission differently. First, coronal and solar wind models are used to \textit{forecast} the connectivity between a spacecraft and the Sun \textit{prior} to each encounter. Accurate forecasts of this connectivity are vital for coordinating in advance simultaneous remote sensing observations of solar wind and SEP source regions using space and ground-based instruments (\textit{e.g.} \textit{SolO}, \textit{DKIST}, \textit{Hinode}). This allows us to directly link detailed measurements (\textit{e.g.} temperature, density, elemental composition, charge state) made at the solar source to in situ spacecraft measurements. Without model-derived forecasts, remote instruments would not necessarily be aligned with the same region of the Sun that produced the solar wind observed at \textit{PSP}.  The second way that models are used, is \textit{retrospectively} to derive the source regions of the solar wind observed by \textit{PSP}. This type of analysis is best for basic research applications because photospheric field observations to drive the model are available for the entire encounter, as opposed to deriving the spacecraft connectivity for a future time with only present data.  However, in both the case of forecasting and retrospective analysis, we do not have observations of the poles and solar far-side to drive coronal models, the effects of which are discussed in detail throughout this paper. Further, retrospective model solutions can also be validated with the \textit{PSP}-observed solar wind speed and interplanetary magnetic field (IMF) measurements. Similarly, in a retrospective study the model-derived coronal hole areas can be compared to EUV observations when available.   The information obtained from both modeling approaches is critical to interpret in situ observations from missions such as \textit{PSP} that observe the pristine solar wind and aim to further understand how the solar wind and SEPs are generated, released, heated, and accelerated.

In this work, we use the Wang-Sheeley-Arge (WSA) model \citep{ArgePizzo2000,Arge2003a,Arge2004,McGregor2008} driven by Air Force Data Assimilative Photospheric Flux Transport (ADAPT: \citealp{ADAPT_Arge2009,Arge2010,Arge2013,Hickmann2015}) time-dependent photospheric field maps to derive the coronal magnetic field, as well as source regions of the solar wind observed at \textit{PSP} for the spacecraft's first two solar encounters, both of which have a perihelion of 35.6 \(R_\odot\).  Since all our model solutions are derived retrospectively as opposed to forecast prior to the encounter, we validate our results by comparing model-derived solar wind speed and magnetic polarity with that observed by the Solar Wind Electrons, Alphas, and Protons (\textit{SWEAP}: \citealp{Kasper2016}) and the Fields experiment (\textit{FIELDS}: \citealp{Bale2016}) instruments onboard \textit{PSP}.   This allows us to provide the most realistic model derived coronal field and solar wind source regions.  When modeling results are very reliable, we present time series of expansion factor, coronal hole boundary distance, and the photospheric magnetic field strength for the field lines that are the source regions of the \textit{PSP}-observed solar wind.  Additionally, we highlight a new and useful capability of WSA developed for the \textit{PSP} mission, in which we quantify the spacecraft's separation from the HCS. We discuss how these parameters can be used together to confirm where the solar wind originated from.  The primary objectives of this paper are to:

\begin{enumerate}[i)]
  \item Identify the source regions of the \textit{PSP}-observed solar wind for the purpose of interpreting how the solar wind was formed.
  \item Discuss specific aspects of each encounter that can only be understood through the use of a model.
   \item Address periods when ADAPT-WSA is unable to reproduced the observed solar wind and determine why.
\end{enumerate}


This paper is structured as follows. Section \ref{sec:2_Model} highlights the unique capabilities of the ADAPT-WSA model compared to other modeling approaches. It also outlines model input parameters specific to this study.  The results for encounters 1 and 2 are presented and discussed in Sections \ref{sec:3_PSP-1-Results} and \ref{sec:4_PSP-2-Results} respectively.  In Section~\ref{sec:5_Conclusion}, we conclude by emphasizing the specific ways that coronal modeling uniquely enables the scientific interpretation and understanding of how the observed solar wind was formed for both encounters, and we discuss the limitations of our modeling.  Our work is summarized in the final section, where we also highlight recent studies that our modeling has supported.\\

\section{The ADAPT-WSA model} \label{sec:2_Model}

The WSA model is a combined empirical and physics-based model of the corona and solar wind. It derives the global coronal magnetic field using a coupled set of magnetostatic potential-field source surface (PFSS) models \citep{Shatten1969,Altschuler1969,WangSheeley1992}.   The model provides the solar wind solution anywhere in the inner heliosphere as well as the source regions of the solar wind. WSA is most recognized as being the first operational model used to forecasting the solar wind at 1 au \citep{Pizzo2011}. As such, it is routinely used along with other coronal models to derive the source regions of the solar wind observed at \textit{Parker Solar Probe}. All model solutions generated in this study were derived with WSA version 5.3.

\subsection{Advantages of using ADAPT-WSA} \label{sec:2.1_advantages}

There are several advantages to using WSA vs. other approaches, some of which are discussed here.  First, while most potential field models only derive the coronal field out to $\sim$2.5$\pm$0.5 \(R_\odot\) (which can prematurely force the field to be radial), WSA additionally derives the outer coronal field with a second potential field type model, the Schatten Current Sheet (SCS) model. Using the traditional PFSS inner coronal solution as input, the SCS model allows WSA to provide a more realistic magnetic field topology of the upper corona (\textit{e.g.} from 2.5\,--\,21.5 \(R_\odot\), \citealp{WangSheeley1995,Arge2004,McGregor2008}). 

Second, in this study WSA is driven by photospheric field maps derived with the ADAPT model.  ADAPT ingests new magnetograms and assimilates them into a synoptic map that represents the Sun \textit{synchronically} (\textit{i.e.} at one moment in time). This is only possible by using flux transport models (\textit{e.g.} \citealp{WordenHarvey2000,SchrijverDeRosa2003}), because they account both for solar time-dependent phenomena (\textit{e.g.} differential rotation, meridional and supergranulation flows), and model and data uncertainties resulting from the lack of far-side and polar photospheric field observations.  In addition, ADAPT is an ensemble model which allows it to provide 12 possible states (\textit{i.e.} realizations) of the solar surface magnetic field, representing the best estimate of the range of possible global photospheric flux distribution solutions at one moment in time. When coupled with ADAPT, WSA derives an ensemble of 12 realizations each representing an estimate of the global state of the coronal field and spacecraft connectivity to 1 \(R_\odot\) for a given moment in time.  This set of solutions provides an estimate of the uncertainty in the ADAPT-WSA solution.  The best realization is then determined by comparing model-derived with observed solar wind speed and interplanetary magnetic field (IMF). In this study, we select ADAPT model maps that produce results that best agree with observations.   Without flux transport models like ADAPT, it is only possible to represent the photosphere \textit{diachronically} (\textit{i.e.} traditional Carrington maps), where one photospheric field map is assembled from near-side central meridian observations in sequence over an entire Carrington rotation (CR). These maps are not ideal for deriving the global coronal field and magnetic connectivity because they do not represent one moment in time, but rather are a time-history of how the central meridian evolved over a CR.  For a more complete summary of the ADAPT model vs other methods, see \citealp{Wallace2020}.
 
The third advantage of using WSA is that we use a unique approach to derive the source regions of the observed solar wind. WSA first determines the location of the spacecraft in time as if it were at the same heliographic latitude and longitude, but positioned at the outer boundary of the coronal magnetic field solution (\textit{i.e.} 5 \(R_\odot\) in this study).  Next, the model identifies the magnetic connectivity between the projected spacecraft position at the outer coronal boundary (referred to as sub-satellite points) and 1 \(R_\odot\).  The model then uses an empirical velocity relationship \citep{Arge2003b, Arge2004} to calculate the solar wind speed at the sub-satellite points.  This relationship is a function of both flux tube expansion factor (\textit{f$_s$}, \textit{e.g.} \citealp{WangSheeley1990,WangSheeley1992,Wang1996}) and coronal hole boundary distance ($\theta_b$ or DCHB, \textit{e.g.} \citealp{Riley2001,Riley2015}). Both parameters are widely used to empirically derive solar wind speed. Expansion factor has been shown to be inversely related to solar wind speed, and is thought to modulate solar wind speed along open field lines in wave-turbulent driven acceleration theories. On the other hand, $\theta_b$ quantifies the minimum angular separation between a field line footpoint and the nearest coronal hole boundary, and thus is a proxy for magnetic reconnection.  There is increasing evidence that the expansion factor-speed relationship originally defined in \citealt{WangSheeley1990} does not hold for pseudostreamers \citep{RileyLuhmann2012,Riley2015,Wallace2020}.  For this reason, the empirical velocity relationship used in WSA has been tuned to rely heavily on $\theta_b$ when the solar wind originates from near the magnetic open-closed boundary.  

Once the solar wind speed is calculated, WSA propagates the solar wind radially \textit{outward} as individual parcels (\textit{i.e.} one from each endpoint of the sub-satellite field lines) to the observing spacecraft. WSA also uses a simple 1-D modified kinematic model to account for stream interactions which prevents fast streams from bypassing slow ones \citep{Arge2004}.  The model provides the time-of-arrival of each solar wind parcel at the spacecraft, the model-derived solar wind speed, in addition to the interplanetary magnetic field (IMF).  Our method is uniquely different than approaches that use a ``backmapping'' methodology.  The backmapping technique assumes a Parker Spiral and traces the observed solar wind from a spacecraft back to the source surface (SS) of a PFSS-derived coronal field by assuming an average speed for the solar wind (typically 400 km s$^{-1}$), or by using the observed speed if the mapping is done retrospectively.  If the backmapping is done restrospectively, then the in situ observed $B_r$ can be compared with the model-derived polarity at the source region. If the source region and in situ observed polarities do not agree, one can adjust the SS height to obtain an agreement between these two quantities. This can make the model-derived coronal hole areas unrealistic by opening/closing too much magnetic flux, especially if there are not remote coronal observations available to compare with. On the other hand, WSA's propagation-based approach requires solar wind speed to be derived for individual parcels, providing another parameter other than $B_r$ to validate the model solution with observations. Also, WSA's ability to account for stream interactions (SIRs), and to derive the coronal field farther out than the SS allows for the model to represent the global coronal magnetic field topology and solar wind outflow more accurately. However, both WSA and backmapping approaches have been shown to produce similar results when SIRs are not present \citep{Kahler2016}.

Fourth, PFSS-based models such as WSA are relatively efficient computationally compared to magnetohydrodynamic (MHD) codes. This is especially useful since ADAPT uses 12 solutions of the photospheric field for a given moment in time to drive WSA.  MHD models make use of more advanced physics to derive the coronal field, yet they have been shown to produce relatively similar coronal topologies to PFSS models \citep{MHDvsPFSS-Riley2006}. Additionally, while MHD and force free models can more accurately determine the coronal magnetic field configuration in non-potential regions such as active regions, the input boundary conditions in these regions are usually the most suspect.  This is because most photospheric field maps make use of the line-of-sight magnetic field measurements along with the radial field assumption, which is known to be a very poor assumption in active regions. It is therefore unclear if using more advanced models improves magnetic connectivity determinations over that of the PFSS model, given the large uncertainty in the input boundary conditions. While static PFSS-based models have their limitations, using an ensemble of synchronic photospheric field maps (such as those provide by ADAPT) to drive WSA both minimizes and helps to quantify the uncertainty in the model solutions. 

While no model is perfect, the capabilities discussed above make ADAPT-WSA an extremely useful and relatively rigorous tool for determining the source locations (along with their corresponding uncertainties) of \textit{PSP} in situ solar wind measurements.

\subsection{Limitations and Uncertainty}

Despite the advantages of using ADAPT-WSA, there are limitations and uncertainties with our modeling approach.  It is important to note that in this work we derive the coronal magnetic field at 2$^\circ$ resolution (\textit{i.e.} grid cells of $\sim$24,000 in diameter at the equator, roughly equivalent to a small supergranule), which sets the spatial resolution limits for deriving the solar wind source regions. Similarly, WSA does not predict temporal variations in the solar wind on scales less than one hour in situ. 

Further, potential field based models have their own inherent uncertainties and limitations.  These models are magnetostatic because they derive the coronal magnetic field based on the assumption that the corona is nearly potential.  This allows for a ``source surface'' (an artificial construct) to be imposed on the coronal solution, forcing the field to be open and radial beyond some height (typically  2.5 \(R_\odot\)). This height is a free parameter that when raised, results in smaller CHs and less total open area.  The opposite holds true when the SS is lowered. Several studies also suggest that the SS height is dependent on solar cycle, as well as input magnetogram data (\citealp{Lee2011}; \citealp{Arden2014}; \citealp{Nikolic2019}). Our justification for the SS heights selected for each encounter are described in Section~\ref{sec:2.3_model_params_for_enc_1_2}.

Similarly, since PFSS models are magnetostatic, a time-dependent energy equation (such as that used in an MHD model) is not required to generate the coronal magnetic field. For this reason, it is not possible for PFSS models to capture the Sun's time dependent phenomena associated with the opening and closing of magnetic flux (\textit{e.g.,} magnetic reconnection, transient eruptions such as coronal mass ejections). While we can account for time-dependent photospheric phenomena with ADAPT, WSA only derives the magnetic connectivity between an observing spacecraft and model-derived field lines that are open.  Similarly, WSA cannot provide information regarding how long a particular field line has been open.  Therefore, when the model predicts that a spacecraft measured plasma near the open-closed boundary, the two physical scenarios that are possible are 1) the plasma originated from that particular open field line, or 2) the plasma originated on closed field that was recently opened via interchange reconnection, whereas ADAPT-WSA cannot make the distinction between the two possible scenarios. For the purposes of this work, it is not entirely necessary to derive the connectivity with this level of specificity. Also, reliably determining this information is currently beyond the capability of all global coronal models due to the inaccuracies of the input boundary conditions (\textit{i.e.} photospheric field maps). 

Lastly, there is a fair amount of uncertainty in virtually all coronal and solar wind models \citep{MacNeice2009a,MacNeice2009b,MacNeice2011, MacNeice2018,NorquistMeeks2010,Norquist2013,Owens2005,Owens2008b}. This is primarily because of large uncertainties in the coronal solutions, due to the lack of knowledge of the specific mechanism(s) heating the corona, as well as those generating and accelerating the solar wind.  Even more problematic, are the large uncertainties in the global photospheric magnetic field maps, which serve as a key driver to these models. \citet{Posner2021} describe in detail the key issues producing these uncertainties. One major issue is that we do not observe the polar magnetic fields, a crucial component to the global coronal field \citep{ArgePizzo2000,Riley2019}. Similarly, global coronal and solar wind solutions can be significantly impacted by the lack of far-side observations \citep{Arge2013,Cash2015}. This can occur when a new active region emerges on the Sun’s far-side and therefore is not included in the photospheric field map used to drive the model. Coronal solutions can also be adversely impacted when these new active regions rotate onto the visible side of the Sun and ``suddenly'' appear in the maps. However, the effects of far-side emergence are mitigated in retrospective analyses such as what we perform in this study. Once the AR is known to the model, the coronal solution stabilizes after 2\,--\,5 days. Lastly, an especially problematic but common scenario arises when only part of a far side active region is incorporated into a photospheric field map. In such cases, a non-physical monopole occurs in the global map that must be removed to ensure that is divergence free (see \citealp{Jones2020} for how monopoles are removed). 

\subsection{Model input parameters specific to \textit{PSP} Encounters 1 and 2} \label{sec:2.3_model_params_for_enc_1_2}

There are a few model inputs and parameters that are specific to this study.  First, we use ADAPT maps generated with magnetograms from the Global Oscillation Network Group (GONG: \citealt{GONG-Harvey1996}) for both solar encounters, because they provide the best overall agreement between ADAPT-WSA output and \textit{PSP} observations vs. using magnetograms from other sources (\textit{e.g.,} \textit{SDO}/HMI). Our model output is derived with a sequence of ADAPT-GONG photospheric field input maps updated at a daily cadence over the entire encounter. For each encounter, the ADAPT-WSA realization that agrees best with \textit{PSP}-observed solar wind speed and IMF is discussed. Second, we experiment with lowering the source surface height to within $\pm$0.5 \(R_\odot\) of the traditional value (\textit{i.e.,} 2.5 \(R_\odot\)).  For the first encounter, 2.0 \(R_\odot\) produced the best agreement overall between WSA-derived and \textit{PSP}-observed solar wind speed and IMF, and between WSA-derived coronal hole and EUV observations (when available). This value agrees with the SS height chosen by \citet{Badman2020} and \citet{Panasenco2020} to agree best with observations for the entire encounter. For the second encounter, we find that defining the SS at 2.5 \(R_\odot\) produces the best agreement overall between models and observations, in agreement with what \citet{Panasenco2020} concluded for the second encounter. Therefore, the modeling results presented in this paper are derived with the SS at 2.0 \(R_\odot\) for Encounter 1, and 2.5 \(R_\odot\) for Encounter 2.

However, both \citet{Badman2020} and \citet{Panasenco2020} experimented with lowering the SS to less than 2.0 \(R_\odot\) to match the \textit{PFSS}-derived radial field with that observed at \textit{PSP} for specific periods within each encounter.  \citet{Panasenco2020} also explored what a non-spherical SS may have looked like for each encounter.  We do not investigate lowering the SS to below 2.0 \(R_\odot\), because the primary goal of our work is slightly different than previous works.  Our goal is to provide the community with detailed information on the source regions of the \textit{PSP}-observed solar wind in order to enable scientific interpretation of \textit{PSP} observations. Therefore, lowering the source surface far below the traditional value (\textit{i.e.} 2.5 \(R_\odot\)) would have to be validated with both $B_r$ \textit{and} remote observations of coronal holes to ensure our model output is realistic. Since these two encounters occur mainly on the far-side, the latter is not possible. Although it is possible to obtain agreement between model-derived and spacecraft-observed $B_r$ by lowering the SS, it could be for a non-physical reason.  In fact, \citet{Nikolic2019} tested various SS heights for PFSS coronal field solutions derived with GONG synoptic maps from 2006\,--\,2018, and found that lowering the SS during solar minimum produced artificial low-latitude coronal holes. They concluded that for solar minimum, in order for the PFSS-derived $B_r$ to agree with observations, the GONG polar fields should be increased, rather than lowering the SS much beyond the traditional value.\\

\section{First solar encounter} \label{sec:3_PSP-1-Results}

\subsection{Results} \label{sec:3.1_Enc1_Results}


\begin{figure}[h!]
\vspace{-0.5cm}
\begin{center}
\includegraphics[width=\textwidth,trim={5.5cm 2.2cm 2cm 0.5cm},clip]{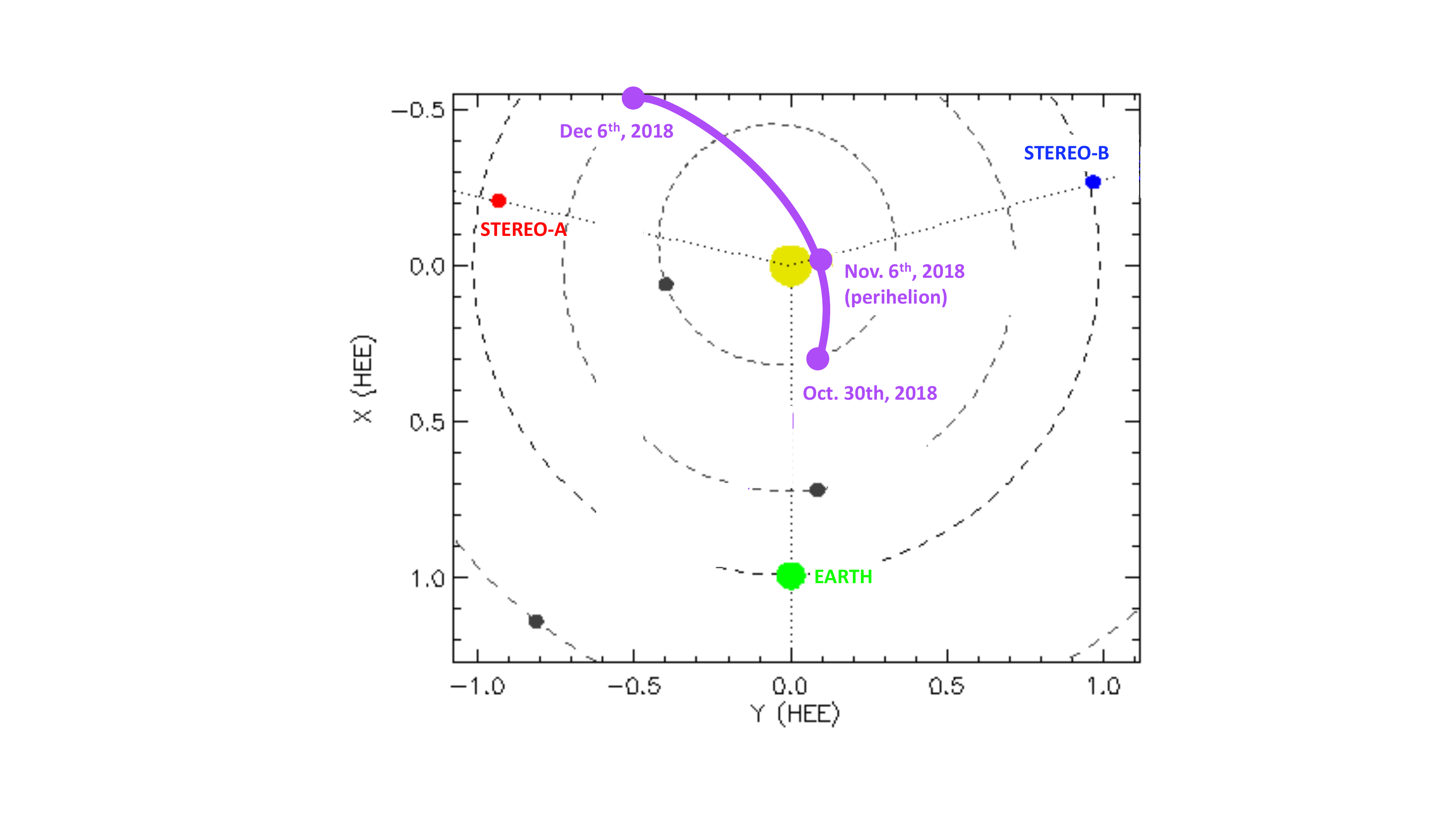} 
\end{center}
\vspace*{-0.2cm}
\caption{ \footnotesize \textit{PSP}'s position and trajectory (purple) for its first solar encounter, overplotted onto the inner solar system in heliographic Earth ecliptic (HEE) coordinates (edited from original figure provided by NASA/GSFC). The location of Earth, \textit{STEREO-A}, and \textit{STEREO-B} are marked in green, red, and blue, respectively, while Mercury, Venus, and Mars are labeled with black circles. }\label{Fig_1:Enc1_orbit}
\end{figure}




\begin{figure}[t!]
\begin{center}
\includegraphics[height=.85\textheight,keepaspectratio]{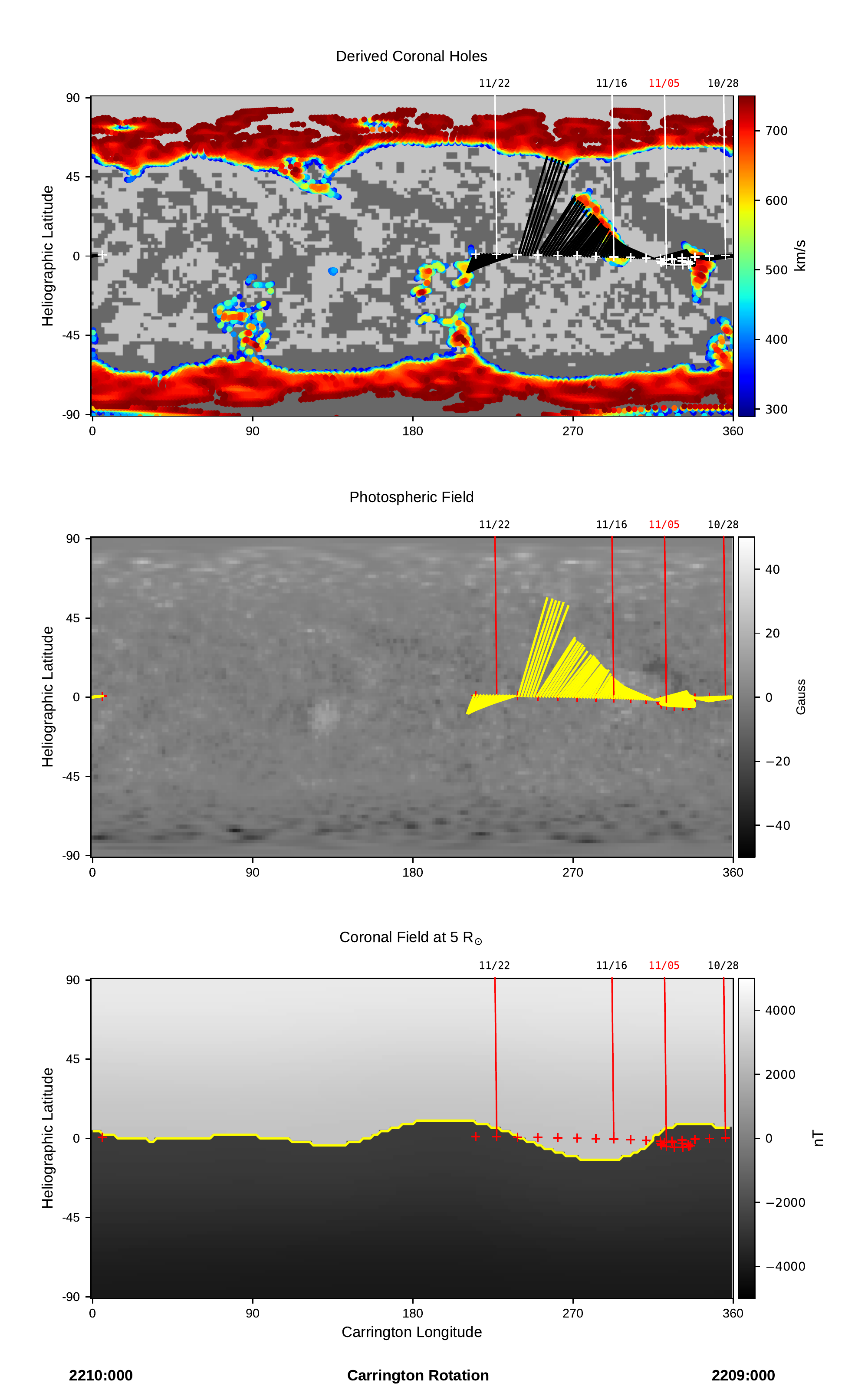}

\end{center}
\caption{ \footnotesize ADAPT-WSA model output for CR 2209\,--\,2210 (October\,--\,November 2018) summarizing the first \textit{PSP} solar encounter. White (a) or red (b,c) tick-marks label the sub-satellite points, representing the heliographic location of \textit{PSP} in time mapped to 5 \(R_\odot\) with dates labeled above in red.  Black (a) or yellow (b) straight lines connect the projection of \textit{PSP}'s location at 5 \(R_\odot\) to the solar wind source region at 1 \(R_\odot\). \textbf{a) (top)} WSA-derived open field at 1 \(R_\odot\) for 5 November 2018 00:00:00 UTC (near perihelion) with model-derived solar wind speed in colorscale.  The field polarity at the photosphere is indicated by the light/dark (positive/negative) gray contours.  \textbf{b) (middle)}  Synchronic ADAPT-GONG photospheric field (Gauss) for 5 November 2018 00:00:00 UTC, which reflects the timestamp of the last magnetogram assimilated into this map. \textbf{c) (bottom)} WSA-derived coronal field at 5 \(R_\odot\).  Yellow contour marks the model-derived HCS. }\label{Fig_2:WSA_Enc1}
\end{figure}




\begin{figure}[t!]
\centering
\includegraphics[height=.83\textheight]{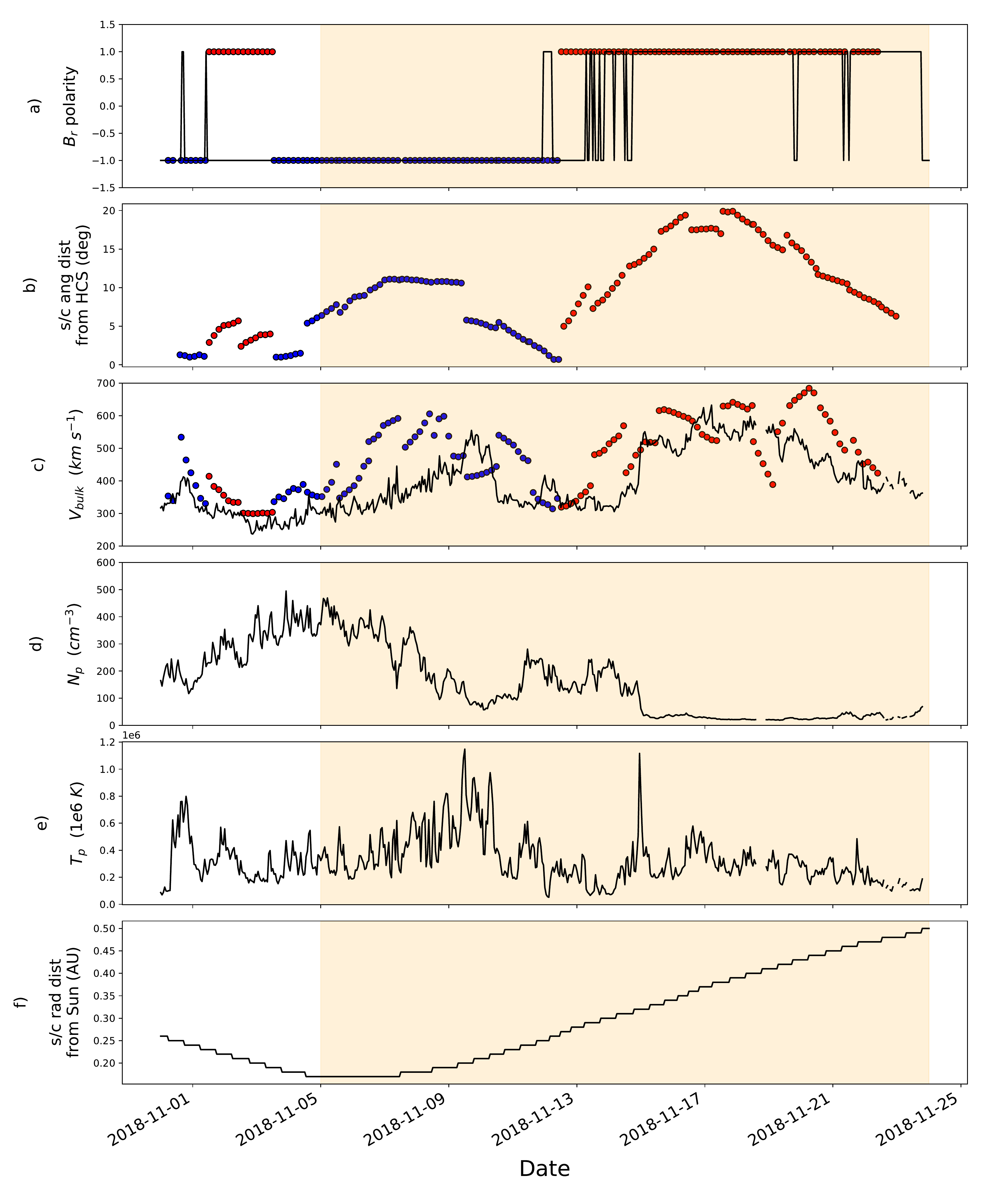}  
\caption{ \footnotesize Hourly-averaged \textit{PSP} data (black) with ADAPT-WSA model output (color) overplotted when available from 30 October\,--\,24 November 2018.  The red and blue colored points represent individual model-derived solar wind parcels (red/positive, blue/negative) that are observed at \textit{PSP}.  The portions of this figure highlighted in orange are periods when the spacecraft was on the far-side of the Sun. \textbf{a)} $B_{r}$ polarity. \textbf{b)}  Spacecraft angular separation from the HCS (degrees). \textbf{c)} Solar wind speed (km s$^{-1}$). \textbf{d)} Proton number density (cm$^{-3}$). \textbf{e)} Proton temperature (1e6 K). \textbf{f)} Spacecraft radial distance from the Sun (au).}\label{Fig_3:PSP_data_with_WSA_1}
\end{figure}



Figure~\ref{Fig_1:Enc1_orbit} shows \textit{PSP}'s orbital path during its first solar encounter.  The encounter began on 30 October 2018 at $\sim$0.25 au (1 au $\sim$ 215 \(R_\odot\)). \textit{PSP} co-rotated with the Sun from $\sim$ 31 October\,--\,13 November 2018, during which the spacecraft approached its first perihelion on 6 November 2018 at 0.16 au (35.6 \(R_\odot\)). The FIELDS and SWEAP instruments remained on until early December 2018 when \textit{PSP} reached $\sim$0.7 au; however, SWEAP data are increasingly unavailable beyond $\sim$24 November, 2018.  This work is focused on identifying and interpreting the evolution of the solar wind source regions of\textit{PSP} observations from 30 October\,--\,24 November 2018. Figure~\ref{Fig_1:Enc1_orbit} also illustrates an important caveat for deriving the solar wind source locations with a model --- most of this encounter occurs on the Sun's far-side, where the photospheric field (a necessary input for coronal models) is not observed. Thus, any \textit{new} far-side flux emergence is not accounted for in our model results until this activity becomes observable from the near-side. However, our results from the first \textit{PSP} encounter will demonstrate that we are able to account for far-side \textit{evolution} (\textit{i.e.} as opposed to new emergence) through a flux-transport model like ADAPT.


\begin{figure}[t!]
\begin{center}
\includegraphics[width=\textwidth,trim={0cm 0.5cm 0cm 0cm},clip]{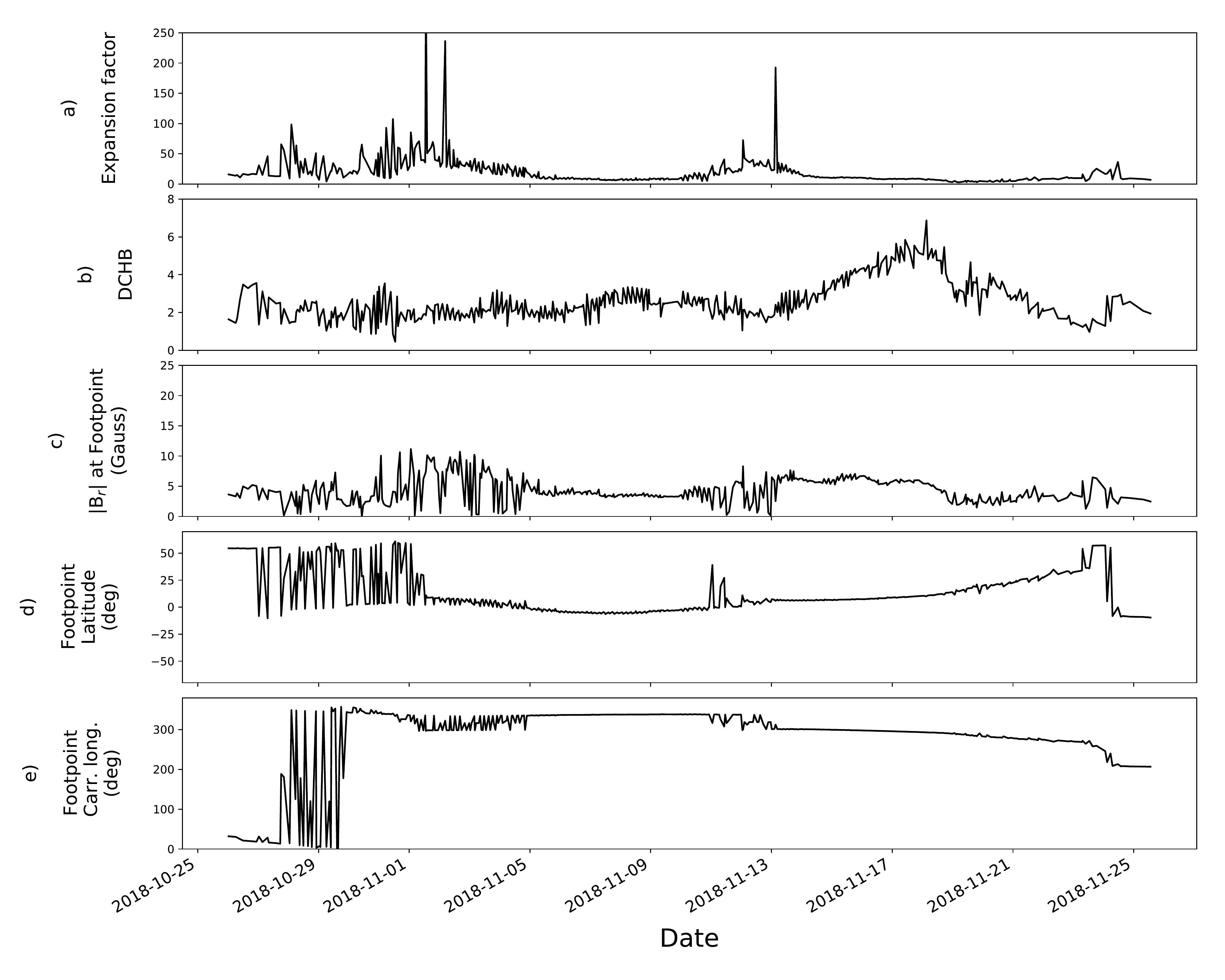} 
\end{center}
\caption{ \footnotesize Hourly-averaged time series of the model-derived (a,b) and observed (c) parameters associated with field lines that are the sources of the \textit{PSP}-observed solar wind.  \textbf{a)} Expansion factor. \textbf{b)} Coronal hole boundary distance. \textbf{c)} $|{B_r}|$ (Gauss). \textbf{d)} Latitude of field line footpoint connectivity (degrees). \textbf{e)} Carrington Longitude of field line footpoint connectivity (degrees).}\label{Fig_4:WSA_params_Enc1}
\end{figure}



Figures~\ref{Fig_2:WSA_Enc1}\,--\,\ref{Fig_4:WSA_params_Enc1} summarize the ADAPT-WSA modeling results for the first \textit{PSP} solar encounter. Figure~\ref{Fig_2:WSA_Enc1} shows the global state of the corona and emerging solar wind at 5 \(R_\odot\) as derived by ADAPT-WSA that best summarizes the period in which \textit{PSP} reaches perihelion and co-rotates with the Sun.  Figure~\ref{Fig_2:WSA_Enc1}a shows the open field line footpoints at 1 \(R_\odot\) (which form the model-derived CHs), with the model-derived solar wind speed in colorscale. The black lines connected to the white cross-hairs (\textit{i.e.} sub-satellite points) and timestamps reveal the precise source regions of the \textit{PSP}-observed solar wind. These dates correspond to when the solar wind \textit{left} the Sun, as opposed to when it \textit{arrived} at \textit{PSP} $\sim$1\,--\,2 days later. Figure~\ref{Fig_2:WSA_Enc1}b shows the same connectivity to the solar surface as in Figure~\ref{Fig_2:WSA_Enc1}a, but overlaid onto the ADAPT-GONG photospheric field map for 5 November 2018.  Figure~\ref{Fig_2:WSA_Enc1}c shows the model-derived global coronal field at 5 \(R_\odot\), with the HCS outlined in yellow.  The \textit{PSP} sub-satellite points are marked in red.  The sub-satellite points do not appear in a uniform fashion as they do for 1 au spacecraft since \textit{PSP} has an elliptical orbit and has periods when it co-rotates with the Sun.  The co-rotation period is identifiable in Fig.~\ref{Fig_2:WSA_Enc1} by the apparent retrograde loop formed by the sub-satellite points. This occurs because the spacecraft does not \textit{precisely} co-rotate with the Sun during this entire time.  

Figure~\ref{Fig_3:PSP_data_with_WSA_1} shows hourly-averaged \textit{PSP} data from 30 October\,--\,24 November 2018.  This figure also compares the WSA-derived polarity of $B_{r}$ and solar wind speed (overplotted in red and blue indicating outward and inward polarity respectively) with that observed at \textit{PSP} to validate our model results.  Panel b of Figure~\ref{Fig_3:PSP_data_with_WSA_1} is a new WSA data product that quantifies \textit{PSP}'s location relative to the HCS (as shown in Fig.~\ref{Fig_2:WSA_Enc1}c), revealing periods when the observed solar wind originates from near to or far from the HCS.  To obtain this information, we calculate the minimum angular separation between the \textit{PSP} sub-satellite points (red cross hairs in Fig.~\ref{Fig_2:WSA_Enc1}c) and the WSA-derived HCS (yellow curve in Fig.~\ref{Fig_2:WSA_Enc1}c). This capability was developed and implemented into WSA specifically for \textit{PSP}, however; this quantity can be calculated for any other satellite incorporated into the model (including \textit{Helios}, see \citealt{DiMatteo2019}).  Lastly, the periods shaded in orange in Fig.~\ref{Fig_3:PSP_data_with_WSA_1} indicate when \textit{PSP} observed solar wind originating from the far-side of the Sun, when new photospheric field observations were not available. Thus, most of this encounter is strictly determined by old observations and flux transport modeling.  

Figure~\ref{Fig_4:WSA_params_Enc1} shows the WSA-derived field line footpoints locations in time that are the sources of the \textit{PSP}-observed solar wind for Encounter 1 (panels d and e), as well as hourly averages of expansion factor ($f_s$), coronal hole boundary distance (DCHB or $\theta_b$), and the magnitude of the ADAPT-GONG observed radial photospheric field component ($|B_{r}|$). It is important to note that the connectivity shown in Figure~\ref{Fig_2:WSA_Enc1} does not precisely align with the parameters derived for individual field lines as shown in  Figs.~\ref{Fig_3:PSP_data_with_WSA_1} and ~\ref{Fig_4:WSA_params_Enc1}.  This is because Figure~\ref{Fig_2:WSA_Enc1} depicts the coronal field at one moment in time (\textit{i.e.} near perihelion in this case), whereas  Figures~\ref{Fig_3:PSP_data_with_WSA_1} \&~\ref{Fig_4:WSA_params_Enc1} are derived with a sequence of photospheric field observations that are updated at a daily cadence over the entire encounter.



\subsection{Encounter 1 Discussion} \label{sec3.2:enc1_discussion}



At the beginning of the encounter (30 October \,--\, 4 November 2018), the WSA-derived HCS is nearly parallel with the ecliptic plane and near the solar equator as seen in Fig.~\ref{Fig_2:WSA_Enc1}c.  This is especially common in periods of low solar activity when there are little to know active regions, which allows for large stable polar CHs form. The observed photospheric magnetic field (Fig.~\ref{Fig_2:WSA_Enc1}b) further supports the model-derived HCS as there are no large active regions, and the \textit{$B_{r}$} associated with the source locations of the \textit{PSP}-observed solar wind is weak QS field (see Fig.~\ref{Fig_4:WSA_params_Enc1}c). During this period, \textit{PSP} skims the HCS and remains within 0.5\,--\,5.0$^\circ$ of the HCS as shown in Fig.~\ref{Fig_3:PSP_data_with_WSA_1}b. This is further supported by the high expansion factors ($f_s>$ 50) derived by the model in Fig.~\ref{Fig_4:WSA_params_Enc1}a from 30 October \,--\, 4 November that are typical of those derived near helmet streamers (which are associated with the HCS). Similarly, during this period the model-derived IMF polarity changes from negative to positive, and then back to negative (Fig.~\ref{Fig_3:PSP_data_with_WSA_1}a) coinciding with the spacecraft being within a few degrees of the HCS (Fig.~\ref{Fig_3:PSP_data_with_WSA_1}c).  These polarity reversals derived by the model are expected when the observed solar wind originates from the HCS, as there is increased uncertainty in the source region of the solar wind at 1 \(R_\odot\).  This is because at 2$^\circ$ resolution, a difference of a few degrees from the HCS (\textit{i.e.} 1\,--\,2 grid cells) could place the spacecraft on one side of the HCS or the other.   Nevertheless, observations confirm (black line in Fig.~\ref{Fig_3:PSP_data_with_WSA_1}a) that for the beginning of the first encounter the \textit{PSP}-observed solar wind is directed inward  (\textit{i.e.,} negative polarity). During this period just prior to perihelion, our modeling results show that \textit{PSP} observes solar wind from near the HCS originating from the open-closed boundary of a mid-latitude CH that extends upward from the southern polar CH. This CH is shown in Fig.~\ref{Fig_2:WSA_Enc1}a as it existed near perihelion (5 November, 2018), at approximately 340$^\circ$ Carrington longitude (CL). For this portion of the encounter, the WSA-derived IMF and solar wind speed agree well with that observed at \textit{PSP} (Fig.~\ref{Fig_3:PSP_data_with_WSA_1}a,c), giving us confidence in our model results.  The observed solar wind speed is slow ($v<$ 400 km s$^{-1}$) with an average speed of $\sim$320 km s$^{-1}$.   The density during this period increases as \textit{PSP} travels closer to the Sun, which is expected since density is dependent on the spacecrafts radial distance from the Sun (\textit{i.e.} $N_p$$\sim$$r^{-2}$).

\begin{figure}[t!]
\begin{center}
\begin{interactive}{animation}{Figure_5_video.mp4}
\includegraphics[width=\textwidth,trim={2cm 1cm 2cm 1.3cm},clip]{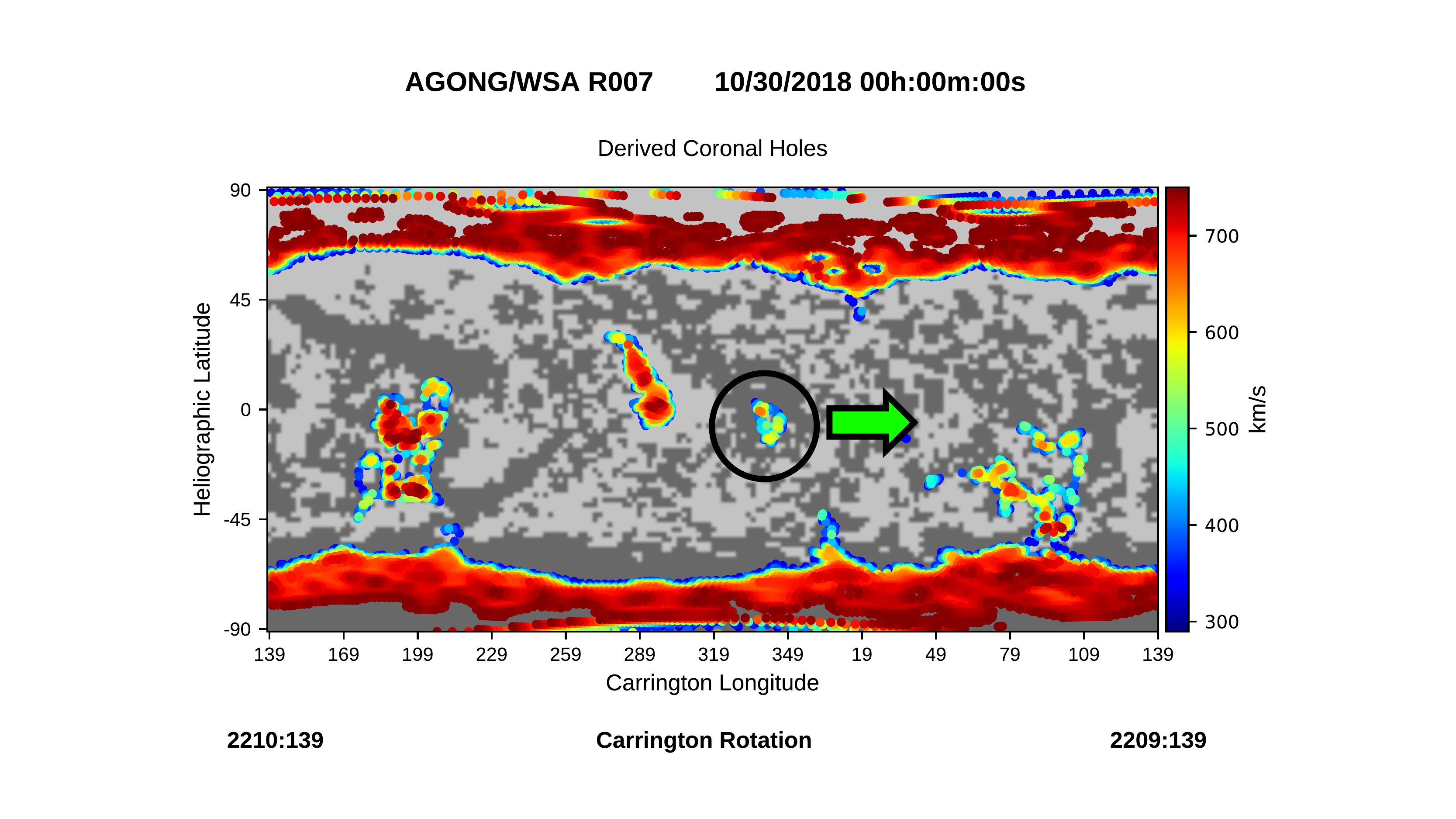} 

\caption{ \footnotesize Animated figure showing the ADAPT-WSA coronal solution evolving in time at a daily cadence from 30 October \,--\, 13 November 2018. The static version of this figure shows the first frame of the animation, with the source region of the \textit{PSP}-observed solar wind at the beginning of its co-rotation with the Sun circled in black.  When viewing the animation, this coronal hole will move to the right as it evolves in time (indicated by green arrow in static version).  The coronal hole increases in surface area from 01\,--\,08 November, 2018, before it starts to close down on 09 November.}\label{Fig_5:movie}
\end{interactive}
\end{center}
\end{figure}

\textit{PSP} continues to observe solar wind that originated from this mid-latitude CH for nearly the entire time that it co-rotates with the Sun ($\sim$30 October \,--\, 11 November).  This can be seen by the spacecraft connectivity to this CH in Fig.~\ref{Fig_2:WSA_Enc1}a (\textit{i.e.} loop of white cross hairs around 330$^\circ$ Carrington longitude), as well as in Fig.~\ref{Fig_2:WSA_Enc1}c which shows that \textit{PSP} is embedded within inwardly directed field for this period. The IMF polarity observations in Fig.~\ref{Fig_3:PSP_data_with_WSA_1}a also confirm this.  This $\sim$13-day period when \textit{PSP} co-rotates with the Sun marks the first time that in situ solar wind has ever been measured from the same source region (in this case a mid-latitude CH) as it evolves in time.  As such, it is important to note while \textit{PSP}'s trajectory in time is overplotted on the ADAPT-WSA output in  Fig.~\ref{Fig_2:WSA_Enc1}, the underlying coronal solution only represents one moment in time (\textit{i.e.} near perihelion) and does not capture the evolution of the coronal field. For this reason we include  Figure~\ref{Fig_5:movie}, an animated figure available in the online version of this work which shows the ADAPT-WSA coronal solution evolving in time in a sequence of snapshots at a daily cadence from 30 October \,--\, 13 November 2018. In this video, the coronal hole increases in surface area consistently from 01\,--\,08 November. This increase in coronal hole open flux with time coincides with an increase in both \textit{PSP}'s predicted separation from the HCS (up to $\sim$12.5$^\circ$ as shown in Fig.~\ref{Fig_3:PSP_data_with_WSA_1}b), and the model-derived and observed solar wind speed (Fig.~\ref{Fig_3:PSP_data_with_WSA_1}a,c). Similarly, the expansion factors of the solar wind source region field lines are low (Fig.~\ref{Fig_4:WSA_params_Enc1}a) and the DCHB for these field lines are moderate (Fig.~\ref{Fig_4:WSA_params_Enc1}b), both of which are consistent with values further into a coronal hole, and away from the magnetic open-closed boundary.  Further, Figure~\ref{Fig_3:PSP_data_with_WSA_1} shows hourly-averaged \textit{PSP} SWEAP and FIELDS data for the solar co-rotation period in which PSP remains connected to the mid-latitude coronal hole (as shown in Fig~\ref{Fig_2:WSA_Enc1}a).  The observations of increasing solar wind speed late into 9 November in panel a are consistent with the coronal hole evolution derived by WSA, in which the open area of this coronal hole increases. As the observed solar wind speed increased, the temperature also increases for this period (Fig.~\ref{Fig_3:PSP_data_with_WSA_1}e), and the density decreases (Fig.~\ref{Fig_3:PSP_data_with_WSA_1}d).  Therefore, both the observations and model output indicate that \textit{PSP} observed the solar wind from deeper within this coronal hole as it grew in size. 

\begin{figure}[t!]
\begin{center}
\includegraphics[width=\textwidth,trim={0cm 5.2cm 0cm 3cm},clip]{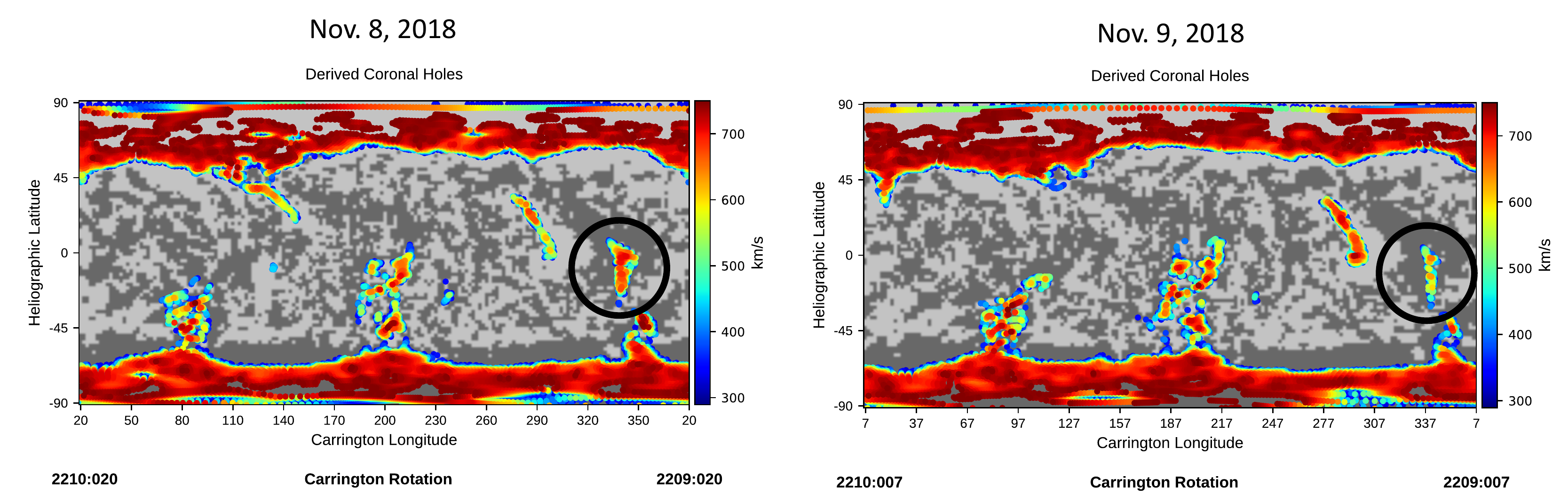} 
\end{center}
\caption{ \footnotesize  WSA-derived coronal holes at 1 \(R_\odot\) for 8 (left) and 9 (right) November 2018 00:00:00 UTC, with the model-derived solar wind speed in colorscale.  The field polarity at the photosphere is indicated by the light/dark (positive/negative) gray contours. The coronal hole circled in black in each plot highlights the mid-latitude coronal hole that \textit{PSP} was connected to while co-rotating with the Sun.}\label{Fig_6:midlatCHevo}
\end{figure}

After 8 November, this coronal hole starts to close down and rapidly shrink over the course of $\sim$1 day. Figure~\ref{Fig_6:midlatCHevo} shows two frames of the animated Figure~\ref{Fig_5:movie} from 8 and 9 Nov. 2018, highlighting the period immediately before and after this mid-latitude coronal hole starts to shrink. The decrease in observed solar wind speed and temperature on 10 November also aligns with when this coronal hole starts to close down as derived by ADAPT-WSA as shown in Fig.~\ref{Fig_6:midlatCHevo}.   Since \textit{PSP} observes the solar far-side during this period, we were only able to compare this model-derived coronal hole with L1 EUV observations on 30 October 2018 when this coronal hole is near disk center.  At this time, the model-derived and observed coronal hole open area agree well on average. 

Further, the model-derived quantities in Fig.~\ref{Fig_4:WSA_params_Enc1} can be used to estimate periods when \textit{PSP} observes the solar wind from precisely the same source region. From approximately 5\,--\,10 November, $f_s$, DCHB, and $B_{r}$ remain nearly constant on average at a time when there is very little change in the location of the field line footpoints.  This suggests that \textit{PSP} remains connected to the same source location within this mid-latitude coronal hole, translating to a set of field lines within $\pm$2$^\circ$ (\textit{i.e.,} 1 grid cell) of one another as derived by the model. This type of measurement has never been made before by any in situ solar wind mission, and the use of a model is the only way to identify the precise window within the solar co-rotation period that this will occur for each \textit{PSP} solar encounter. When \textit{PSP} truly co-rotates with the Sun, any variability observed in situ is due to time-dependent evolution of the solar source region. Outside of solar co-rotation, in situ spacecraft observe a mix of spatial structure and time-dependent evolution as the spacecraft connectivity changes from one source region to another. ADAPT-WSA model output helps to identify when \textit{PSP} observes the solar wind from the same source region over several days, or when the connectivity changes from one source region to another, allowing us to distinguish between spatial and time-dependent variability.  


As \textit{PSP} comes out of solar co-rotation, the spacecraft crosses a well-inclined HCS on $\sim$13 November (Fig.~\ref{Fig_2:WSA_Enc1}c). The WSA-derived polarity changes about a day early when compared to the observations (Fig.~\ref{Fig_3:PSP_data_with_WSA_1}a, 12\,--\,13 November). However, this is not surprising because \textit{PSP} observes a coronal mass ejection (CME) during this period, and a magnetostatic model such as ADAPT-WSA cannot derive time-dependent eruptions. On 11 November at 23:50 UTC, \textit{PSP} observed a slow ($v$ $\sim$ 380 km s$^{-1}$) CME that at the time was the closest to the Sun that a CME had ever been observed in situ. Coincidentally, this eruption was also observed remotely in \textit{STEREO-A}:COR1 and COR2 data. Since this CME was observed remotely, we were able to determine when this event first emerged in the low corona, and then derive the WSA coronal field for that moment in time.  We then identified the source region of this CME as a large helmet streamer formed by the northern polar coronal hole and the inwardly directed mid-latitude coronal hole that PSP co-rotated with. With our modeling, we were able to provide convincing evidence that this CME was a streamer blowout (SBO), which are slow eruptions that often originate in quiet periods from polar crown filaments \citep{VourlidasWebb2018}.  A full analysis of this CME, including the WSA modeling of the source region is shown in \citealp{Korreck2020,Nieves-Chinchilla2020}. 

While crossing the HCS $\sim$13 November, the spacecraft connectivity changes from one mid-latitude coronal hole to another of opposite polarity (\textit{i.e.} positive). This connectivity change is shown in Fig.~\ref{Fig_2:WSA_Enc1}a between $\sim$290\,--\,320$^\circ$ CL during CR 2210.  From 13\,--\,15 November, the observed polarity of $B_{r}$ (black line in Fig.~\ref{Fig_3:PSP_data_with_WSA_1}a) fluctuates from positive to negative, indicative that the spacecraft is still observing wind at the HCS and thus at the boundaries of these mid-latitude coronal holes. Once the spacecraft-observed polarity of $B_{r}$ stabilizes and becomes positive on 15 November, the velocity jumps from 380\,--\,520 km s$^{-1}$, coinciding with a sharp decrease in density. At this same time, the coronal hole boundary distance is increasing (Fig.~\ref{Fig_4:WSA_params_Enc1}b), and the spacecraft is moving away from the HCS (Fig.~\ref{Fig_3:PSP_data_with_WSA_1}b). Therefore, our modeling of the connectivity for this period suggests that this concomitant change in the polarity of $B_{r}$, solar wind speed, and proton number density resulted from the change in connectivity at the solar surface. From 15\,--\,20 November, the spacecraft becomes well-separated from the HCS (up to 20 degrees away as shown in Fig.~\ref{Fig_3:PSP_data_with_WSA_1}b), and observes moderate to fast (500\,--\,610 km s$^{-1}$), low density solar wind (Fig.~\ref{Fig_3:PSP_data_with_WSA_1}c,d)).  The coronal hole boundary distances of the solar wind source region field lines during this period are also high ($\theta_{b} > 4$), with low expansion factors (Fig.~\ref{Fig_4:WSA_params_Enc1}a,b), all of which indicates that \textit{PSP} observed the solar wind from deep inside this mid-latitude coronal hole for $\sim$6 days. 


The ADAPT-WSA model output for Encounter 1 has been compared with several other PFSS and MHD results in \citet{Szabo2020}, all of which agree on average, and derive the same gross structure of the HCS. This is thought to be due to the fact that the Sun was relatively quiet during this encounter, and there was no new significant flux emergence on the far-side.  Our results have been used to identify the source region of the 11 November, 2018 SBO-CME \citep{Korreck2020, Nieves-Chinchilla2020}, and also to identify and characterize small SEP events during the first encounter and understand how these energetic events are formed back at the Sun \citep{Hill2020}. However, it is important to re-emphasize that for most of Encounter 1, \textit{PSP} observes solar wind that originates from the solar far-side, where we do not observe the photospheric magnetic field.  Nevertheless, the WSA-derived and \textit{PSP}-observed IMF and solar wind speed overall track well with each other even despite the lack of photospheric field observations beyond 5 November.   This is because the photospheric field was relatively quiet with no significant new far-side AR emergence, and thus was well-represented by the ADAPT model.

\section{Second solar encounter} \label{sec:4_PSP-2-Results}

\subsection{Results} \label{sec:4.1_Enc2_Results}


\begin{figure}[h!]
\begin{center}
\includegraphics[width=\textwidth,trim={0cm 1.2cm 0cm 0.5cm},clip]{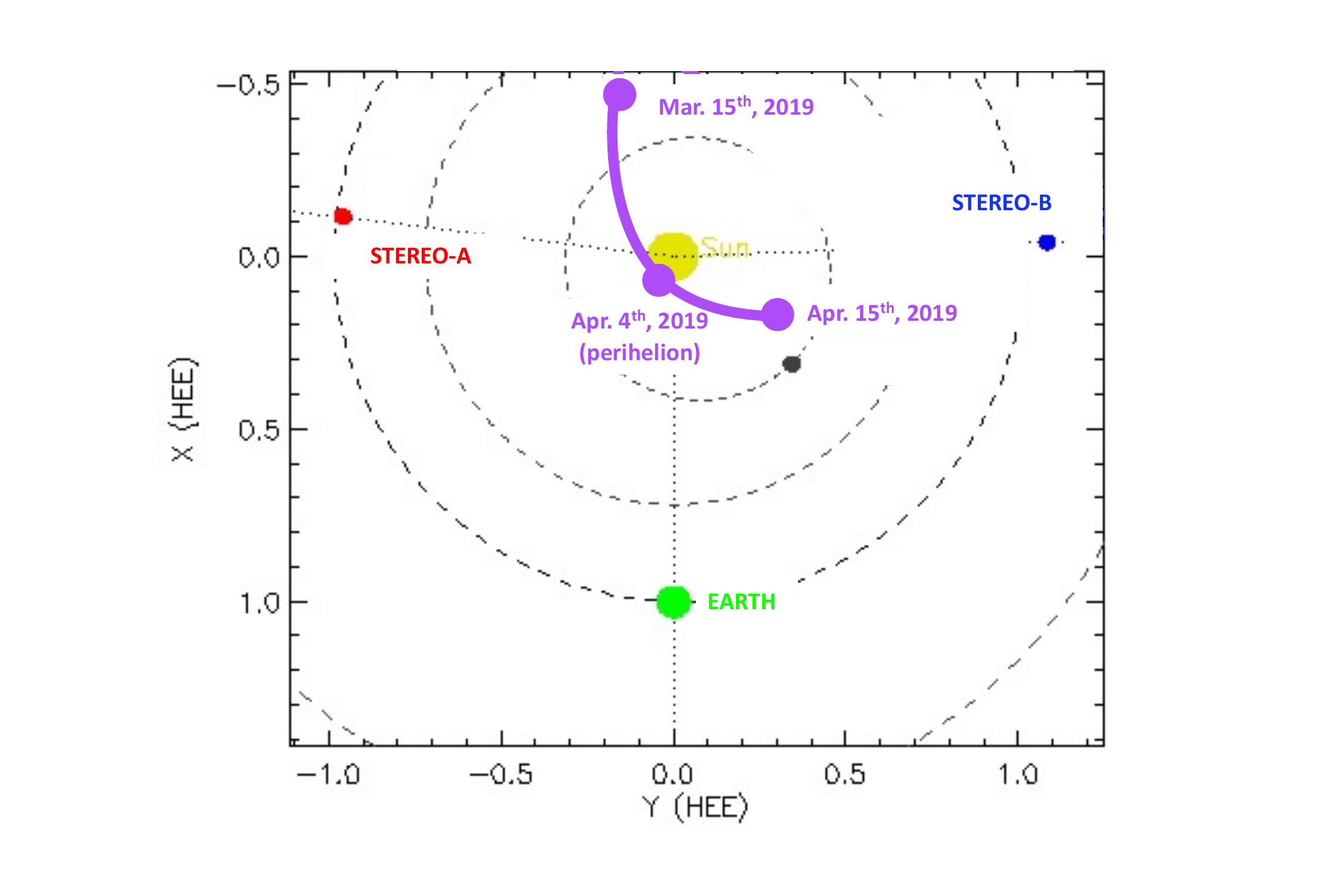} 
\end{center}
\caption{ \footnotesize As in Fig.~\ref{Fig_1:Enc1_orbit}, but for the second \textit{PSP} solar encounter.}\label{Fig_7:Enc2_orbit}
\end{figure}




\begin{figure}[p]
\begin{center}
\includegraphics[height=.85\textheight,keepaspectratio]{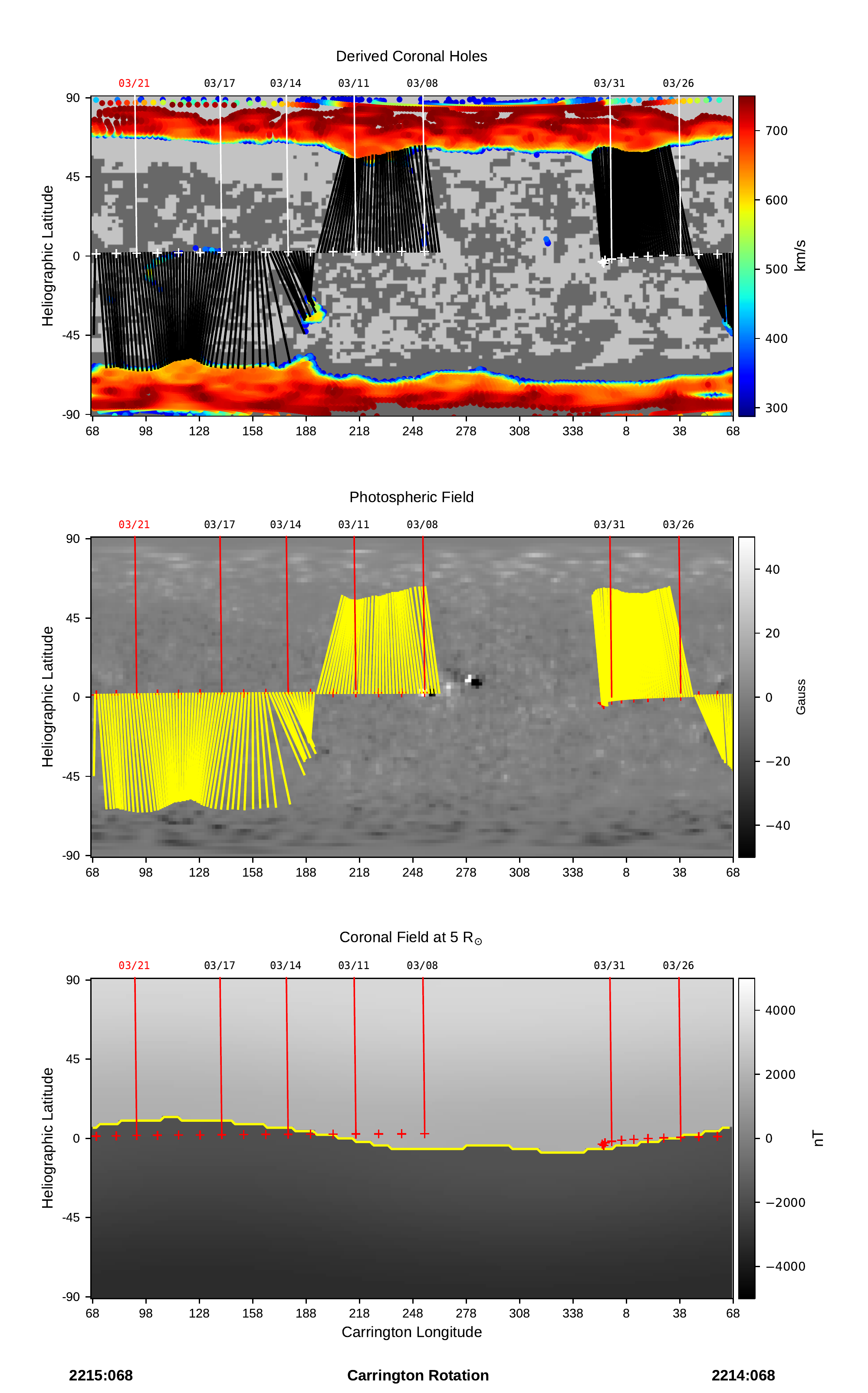} 
\end{center}
\caption{ \footnotesize ADAPT-WSA model output for CR 2214\,--\,2215 (March 2019) summarizing the inbound leg of the second \textit{PSP} solar encounter. White (a) or red (b,c) tick-marks label the sub-satellite points, representing the heliographic location of \textit{PSP} in time mapped to 5 \(R_\odot\) with dates labeled above in red.   Black (a) or yellow (b) straight lines connect the projection of \textit{PSP}'s location at 5 \(R_\odot\) to the solar wind source region at 1 \(R_\odot\). \textbf{a) (top)}  WSA-derived open field at 1 \(R_\odot\) for 21 March 2019 00:00:00 UTC with model-derived solar wind speed in colorscale.  The field polarity at the photosphere is indicated by the light/dark (positive/negative) gray contours. \textbf{b) (middle)}  Synchronic ADAPT-GONG photospheric field (Gauss) for 21 March 2019 00:00:00 UTC, which reflects the timestamp of the last magnetogram assimilated into this map. \textbf{c) (bottom)} WSA-derived coronal field at 5 \(R_\odot\).  Yellow contour marks the model-derived HCS.}\label{Fig_8:WSA_Enc2_inbound}
\end{figure}




\begin{figure}[p]
\begin{center}
\includegraphics[height=.85\textheight,keepaspectratio]{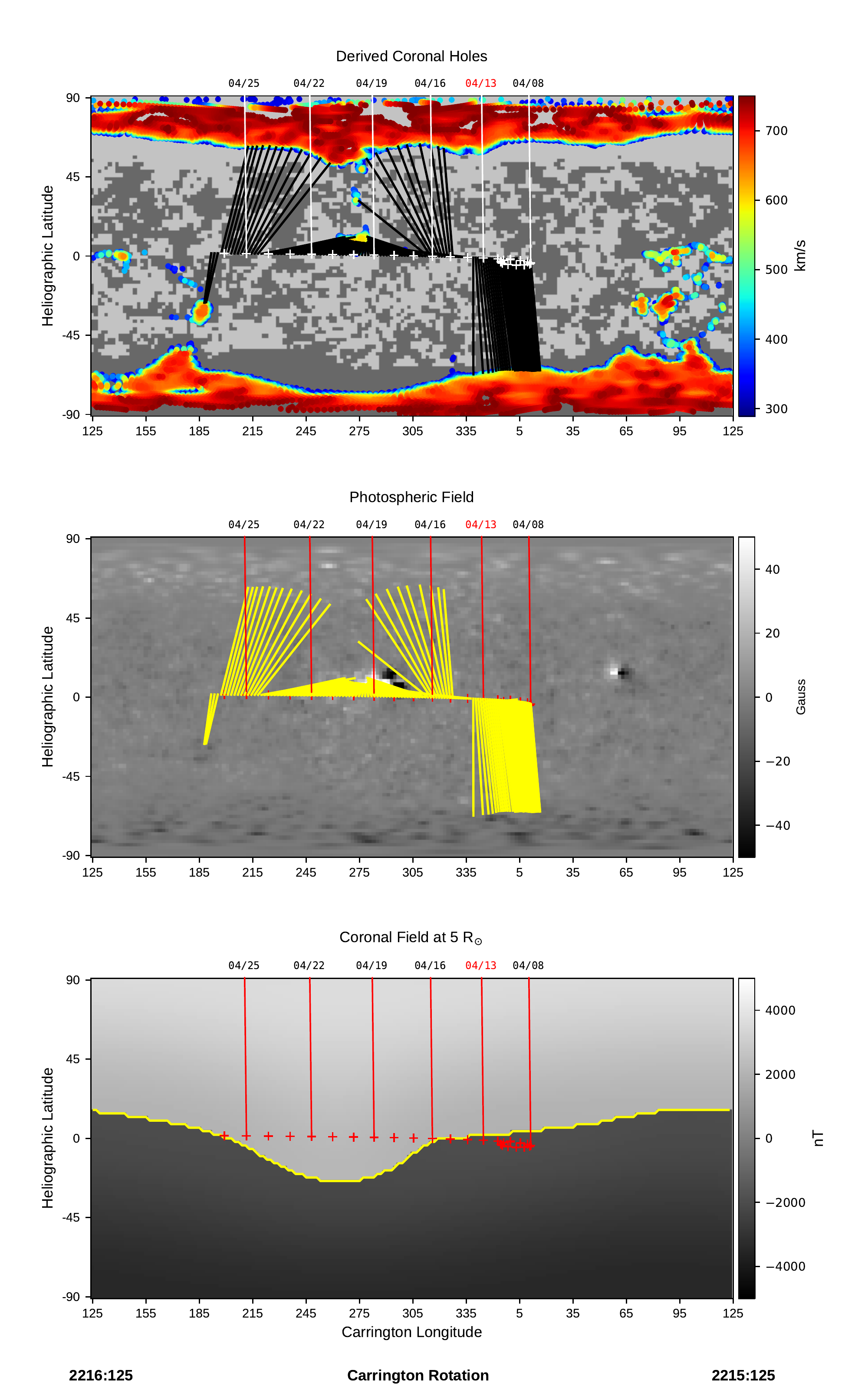} 
\end{center}
\caption{ \footnotesize ADAPT-WSA model output for CR 2215\,--\,2216 (April 2019) summarizing \textit{PSP}'s co-rotation with the Sun in the second encounter. White (a) or red (b,c) tick-marks label the sub-satellite points, representing the heliographic location of \textit{PSP} in time mapped to 5 \(R_\odot\) with dates labeled above in red.   Black (a) or yellow (b) straight lines connect the projection of \textit{PSP}'s location at 5 \(R_\odot\) to the solar wind source region at 1 \(R_\odot\). \textbf{a) (top)}  WSA-derived open field at 1 \(R_\odot\) for 13 April 2019 00:00:00 UTC with model-derived solar wind speed in colorscale.  The field polarity at the photosphere is indicated by the light/dark (positive/negative) gray contours.  \textbf{b) (middle)}  Synchronic ADAPT-GONG photospheric field (Gauss) for 13 April 2019 00:00:00 UTC, which reflects the timestamp of the last magnetogram assimilated into this map. \textbf{c) (bottom)} WSA-derived coronal field at 5 \(R_\odot\).  Yellow contour marks the model-derived HCS.}\label{Fig_9:WSA_Enc2_peri}
\end{figure}




\begin{figure}[t!]
\centering
\includegraphics[height=.84\textheight]{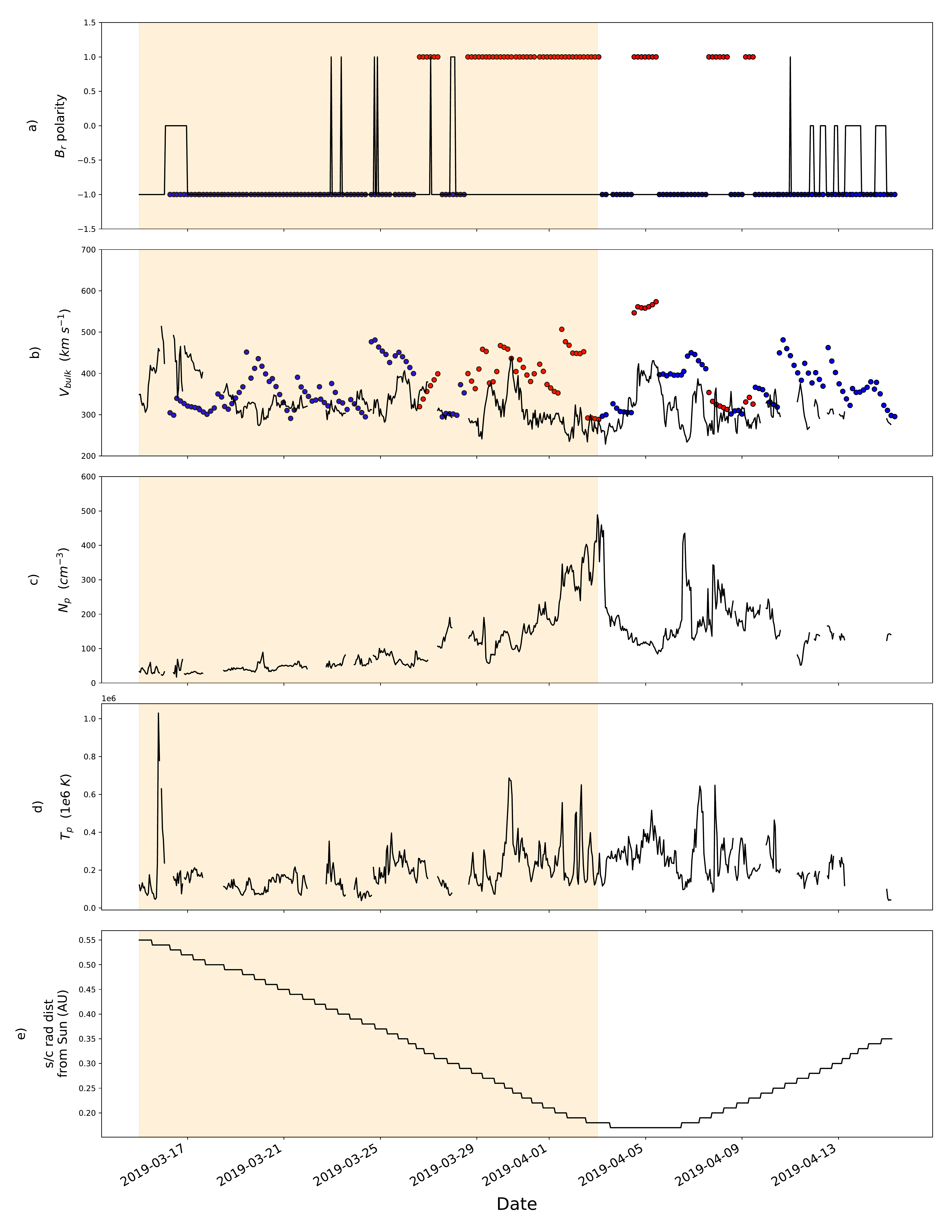} 
\caption{ \footnotesize Hourly-averaged \textit{PSP} data (black) with ADAPT-WSA model output (color) overplotted when available from 15 March\,--\,15 April, 2019.  The red and blue colored points represent individual model-derived solar wind parcels (red/positive, blue/negative) that are observed at \textit{PSP}.  The portions of this figure highlighted in orange are periods when the spacecraft was on the far-side of the Sun. \textbf{a)} $B_{r}$ polarity. \textbf{b)}  Solar wind speed (km s$^{-1}$). \textbf{c)} Proton number density (cm$^{-3}$). \textbf{d)} Proton temperature (1e6 K). \textbf{e)} Spacecraft radial distance from the Sun (au).}\label{Fig_10:PSP_data_with_WSA_2}
\end{figure}




\begin{figure}[t!]
\begin{center}
\includegraphics[width=.9\textwidth]{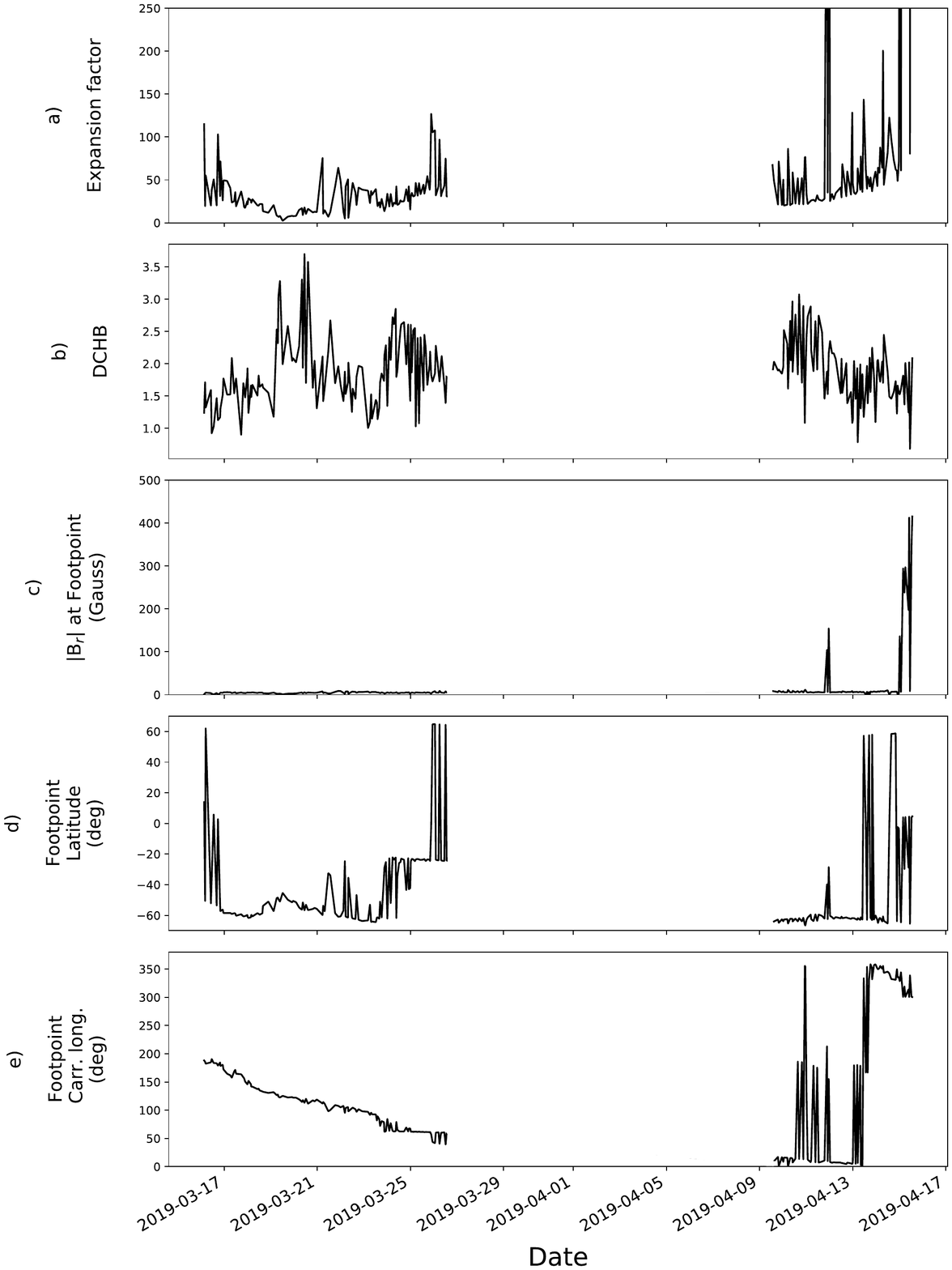} 
\end{center}
\caption{ \footnotesize Hourly-averaged time series of the model-derived (a,b) and observed (c) parameters associated with field lines that are the sources of the \textit{PSP}-observed solar wind. This plot does not contain data from 27 March \,--\, 9 April, 2019 because new far-side active region emergence caused the model output to be unreliable during this period. \textbf{a)} Expansion factor. \textbf{b)} Coronal hole boundary distance. \textbf{c)} $|{B_r}|$ (Gauss). \textbf{d)} Latitude of field line footpoint connectivity (degrees). \textbf{e)} Carrington Longitude of field line footpoint connectivity (degrees).}\label{Fig_11:WSA_params_Enc2}
\end{figure}


Figure~\ref{Fig_7:Enc2_orbit} shows \textit{PSP}'s orbital path during its second solar encounter.  This work discusses the observing period from 15 March \,--\, 15 April, 2019. This encounter began with \textit{PSP} on the solar far-side at a radial distance of 0.55 au from the Sun. The spacecraft started to observe the near-side as it approached a perihelion of 0.16 au (35.7 \(R_\odot\)) on 4 April 2019. By 15 April, the spacecraft was at 0.35 au. As in \textit{PSP}'s first solar encounter, a significant portion of the second encounter occurred while the spacecraft was observing the solar far-side.

Figures~\ref{Fig_8:WSA_Enc2_inbound}\,--\,\ref{Fig_11:WSA_params_Enc2} summarize the ADAPT-WSA modeling results for the second \textit{PSP} solar encounter. Figure~\ref{Fig_8:WSA_Enc2_inbound} shows ADAPT-WSA model output that best depicts the spacecraft connectivity to the solar surface and global coronal magnetic field for the inbound leg of the second encounter (15\,--\,30 March, 2019).  This figure was derived with an ADAPT-GONG photospheric field map from 21 April 2019, as this map produced the most realistic model solution when compared to \textit{PSP} observations of the inbound leg of encounter 2. Figure~\ref{Fig_9:WSA_Enc2_peri} is the same model output as Fig.~\ref{Fig_8:WSA_Enc2_inbound}; however, Fig.~\ref{Fig_9:WSA_Enc2_peri} best summarizes the period in which \textit{PSP} co-rotates with the Sun (30 March \,--\, 12 April, 2019).  After 13 April, 2019, the  model-output shown in Fig.~\ref{Fig_9:WSA_Enc2_peri} is not reliable. This way of representing ADAPT-WSA model output is described in detail in Section~\ref{sec:3.1_Enc1_Results}.

Figure~\ref{Fig_10:PSP_data_with_WSA_2} shows hourly-averaged \textit{PSP} data from 15 March\,--\,15 April 2019.  This figure also compares the WSA-derived polarity of $B_{r}$ and solar wind speed (overplotted in red and blue indicating outward and inward polarity respectively) with that observed at \textit{PSP} to validate our model results.  As in Fig.~\ref{Fig_3:PSP_data_with_WSA_1}, the periods shaded in orange in Fig.~\ref{Fig_10:PSP_data_with_WSA_2} indicate when \textit{PSP} observed solar wind originating from the far-side of the Sun, when new photospheric field observations were not available.  However, unlike Fig.~\ref{Fig_3:PSP_data_with_WSA_1},  Fig.~\ref{Fig_10:PSP_data_with_WSA_2} does not include the spacecraft separation from the HCS due to the reliability of this parameter being suspect for a large portion of this encounter (discussed in Section~\ref{sec:4.2_Enc2_discussion}).

Figure~\ref{Fig_11:WSA_params_Enc2} shows the field line footpoints locations in time that are the sources of the \textit{PSP}-observed solar wind for Encounter 2 (panels d and e), as well as hourly averages of $f_s$, DCHB, and the magnitude of the ADAPT-GONG observed ${B_r}$.  From 27 March \,--\, 9 April, the ADAPT-WSA model output was not reliable enough to derive these parameters (further discussed in Section~\ref{sec:4.2_Enc2_discussion}), and therefore they are not included in this figure. As in Encounter 1, the connectivity shown in Figures~\ref{Fig_8:WSA_Enc2_inbound} and~\ref{Fig_9:WSA_Enc2_peri} does not precisely align with the parameters derived for individual field lines as shown in Figs.~\ref{Fig_10:PSP_data_with_WSA_2} and~\ref{Fig_11:WSA_params_Enc2}.  This is due to the fact that Figures~\ref{Fig_8:WSA_Enc2_inbound} and ~\ref{Fig_9:WSA_Enc2_peri} represent the coronal field at one moment in time during the inbound leg and near perihelion respectively, whereas the model output in  Figures~\ref{Fig_10:PSP_data_with_WSA_2} and~\ref{Fig_11:WSA_params_Enc2} are derived with a sequence of photospheric field observations that are updated at a daily cadence over the entire encounter.


\subsection{Discussion} \label{sec:4.2_Enc2_discussion}


\begin{figure}[t!]
\includegraphics[width=\textwidth,trim={1.7cm 0cm 1.2cm 0cm},clip]{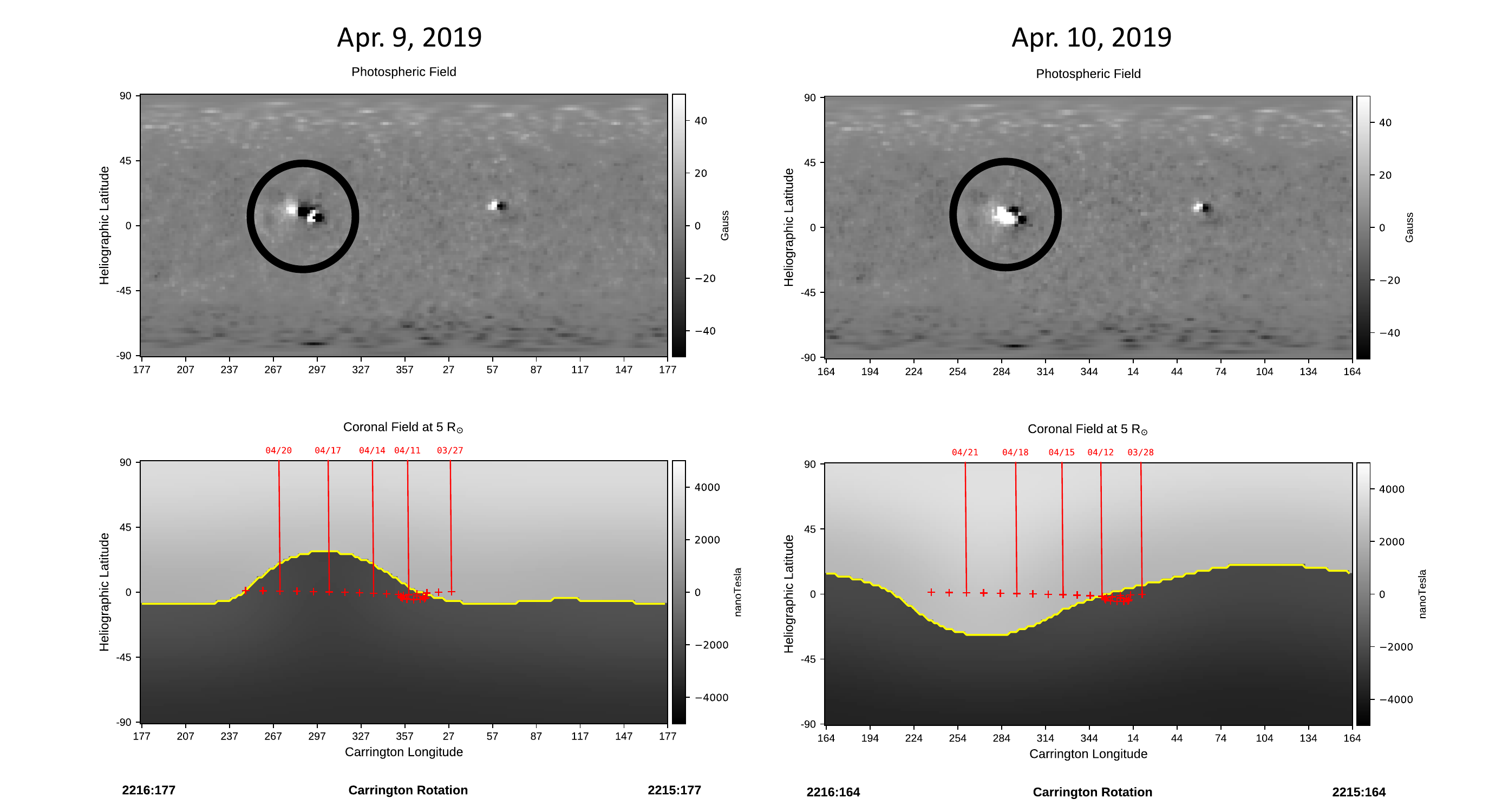} 
\caption{ \footnotesize ADAPT-WSA model output highlighting new far-side active region emergence that rotates onto the east limb on 9 April 2019.  \textbf{a) (top)}  ADAPT-GONG photospheric field map for 9 April (left) and 10 April (right), with the active region circled in black.  \textbf{b) (bottom)} WSA-derived coronal field at 5 \(R_\odot\) for 9 April (left) and 10 April (right).  Yellow contour marks the model-derived HCS. Red tick-marks label the sub-satellite points, representing the heliographic location of \textit{PSP} in time at 5 \(R_\odot\) with dates labeled above in red.}\label{Fig_12:WSA_Apr2019_AR}
\end{figure}




\begin{figure}[t!]
\begin{center}
\includegraphics[width=0.9\textwidth,trim={0cm 0.65cm 0cm 0cm},clip]{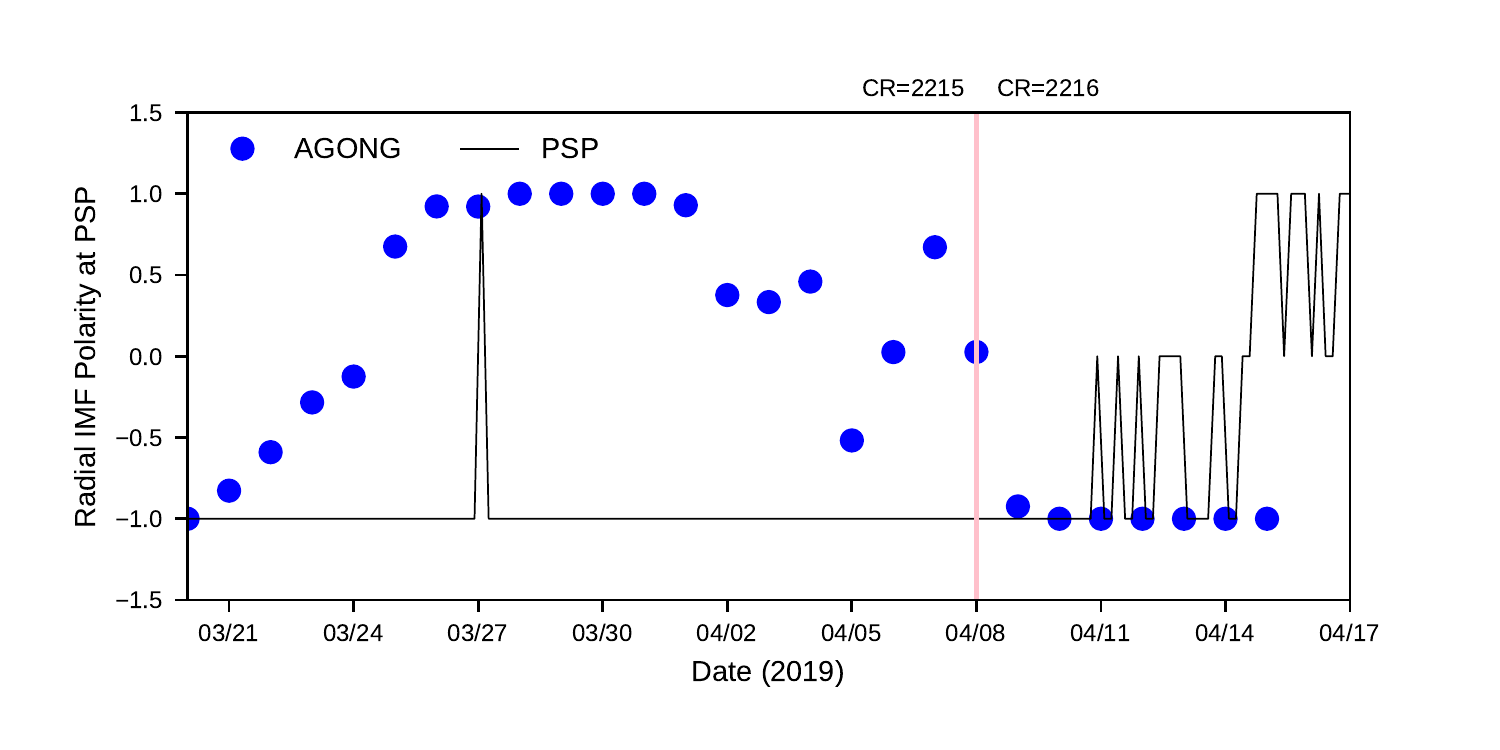} 
\end{center}
\caption{ \footnotesize ADAPT-WSA IMF polarity (blue) derived from sequence of photospheric field observations that are updated at a daily cadence, vs. \textit{PSP} observations (black) for the second solar encounter. Each blue dot represents the average of the ADAPT-WSA IMF polarity over all 12 realizations for individual model-derived solar wind parcels observed at \textit{PSP}.}\label{Fig_13:imf_12avg}
\end{figure}



From 15\,--\,30 March, 2019 (\textit{i.e.} inbound leg of second encounter), the WSA-derived HCS is nearly parallel to the ecliptic plane and near the solar equator ( Fig.~\ref{Fig_8:WSA_Enc2_inbound}c), similar to the first encounter . During this period, \textit{PSP} remains within 5\,--\,7$^\circ$ of the HCS, and observes the solar wind from the boundary of the southern polar coronal hole (Fig.~\ref{Fig_8:WSA_Enc2_inbound}a).  Based on the model-derived coronal hole boundary distance in Fig.~\ref{Fig_11:WSA_params_Enc2}, there are periods in which \textit{PSP} observes the solar wind from slightly deeper inside this coronal hole (\textit{i.e.} periods of 19\,--\,20 March and 24\,--\,26 March).  However, the DCHB is never large enough (\textit{i.e.} $\theta_{b} >$ 4) such that \textit{PSP} observed solar wind from deep inside this coronal hole.  The photospheric magnetic field at the solar wind source regions for this period is weak (Fig.~\ref{Fig_8:WSA_Enc2_inbound}b), with very low values for the magnitude of $B_{r}$ (Fig.~\ref{Fig_11:WSA_params_Enc2}c). Further, on average \textit{PSP} observes slow solar wind during this period (Fig.~\ref{Fig_10:PSP_data_with_WSA_2}b) that correlates reasonably well with the observed temperatures (Fig.~\ref{Fig_10:PSP_data_with_WSA_2}d). The gradual increase in density from 17 March\,--\,3 April, 2019 shown in Fig.~\ref{Fig_10:PSP_data_with_WSA_2}c is due to the density being dependent on the spacecraft radial distance from the Sun ($N_p$$\sim$$r^{-2}$). 

From 30 March \,--\, 12 April, 2019, \textit{PSP} co-rotates with the boundary of the southern polar coronal hole of negative (inward) polarity (Fig.~\ref{Fig_9:WSA_Enc2_peri}a) and remains close to the HCS (Fig.~\ref{Fig_9:WSA_Enc2_peri}c).  The observed solar wind for this period is slow (v $<$ 500 km s$^{-1}$ as shown in Fig.~\ref{Fig_10:PSP_data_with_WSA_2}b).  However, the WSA-derived $B_{r}$ polarity and solar wind speed on average do not agree with \textit{PSP} observations from 29 March \,--\, 9 April, 2019. Figure~\ref{Fig_12:WSA_Apr2019_AR} illustrates why this discrepancy occurs, showing the observed photospheric field (top) and WSA-derived coronal field (bottom) on 9 and 10 April, 2019.  A well-defined pairing of active regions is identified in the black circles in this figure.  One of these active regions emerges on the far-side at some point while \textit{PSP} co-rotates with the Sun.  Since we do not have observations of the far-side,  this active region does not become known to ADAPT until it rotates onto the east limb on 9 April 2019.  The ADAPT-WSA coronal solution solution preceding 9 April looks similar to the solution for 9 April shown in Fig.~\ref{Fig_12:WSA_Apr2019_AR} (bottom left).  However, once the new far-side flux emergence is assimilated into the ADAPT photospheric field map (shown on 10 April in Figure~\ref{Fig_12:WSA_Apr2019_AR}), the ADAPT-WSA solution suddenly changes. In reality, as this AR emerged on the solar far-side (at some time before 9 April), the Sun slowly evolved in time to the new global configuration on 10 April (Figure~\ref{Fig_12:WSA_Apr2019_AR}). However, this evolution cannot be captured by WSA since we do not have observations for when this new flux emergence occurred.  

Figure~\ref{Fig_13:imf_12avg} shows the model-derived polarity of $B_{r}$ in blue as an average of all 12 ADAPT-WSA realizations and that observed at \textit{PSP} in black. The blue dots converge on either positive or negative polarity, denoted by +1 and -1 respectively, for periods when all 12 realizations are mostly in agreement, and fall somewhere in between these two values when the realizations are not in agreement.  For the period in which the model does not know about the new far-side AR emergence, all 12 realizations of ADAPT-WSA output derive the wrong polarity for $B_{r}$ when compared to the \textit{PSP}-observed solar wind.  However, once this AR is assimilated into the ADAPT photospheric field map on 9 April, all 12 realizations then derive the correct polarity observed at the spacecraft.  


\textit{PSP}'s second solar encounter illustrates the importance of having global photospheric field data to drive coronal models. Although the new far-side AR emergence was not the source region of the \textit{PSP}-observed solar wind, it still contributed enough to the total solar magnetic flux to suddenly change the WSA-derived coronal field globally. This new flux emergence also resulted in the model incorrectly deriving the source regions of the \textit{PSP}-observed solar wind for nearly 14 days. For this reason, Figure~\ref{Fig_9:WSA_Enc2_peri} was generated with an ADAPT map from 13 April to allow a few days for this new AR to assimilate into the photospheric field map, thus providing the most realistic summary of the \textit{PSP} co-rotation period with the Sun. 

Despite the difficulties with modeling the second \textit{PSP} solar encounter, we were still able to retrospectively derive the source regions of the \textit{PSP}-observed solar wind once we could incorporate the new far-side AR emergence into our model solution. For the inbound leg and while co-rotating with the Sun, \textit{PSP} observes the solar wind originating from quiet Sun magnetic field at the boundary of the southern polar coronal hole, and remains within $\sim$7$^\circ$ of the HCS. However, during the period in which the far-side AR was unknown to the model (29 March \,--\, 9 April, 2019), we are less confident in the precise model-derived source locations of the \textit{PSP}-observed solar wind. Therefore, we do not discuss WSA-derived parameters as shown in Fig.~\ref{Fig_11:WSA_params_Enc2}) during this period.   



\section{Discussion} \label{sec:5_Conclusion}

To fully address the \textit{PSP} science objectives, it is necessary to identify the specific source locations (\textit{i.e.} which coronal hole or coronal hole boundary) of the solar wind observed at the spacecraft and associated magnetic field at 1 \(R_\odot\) (\textit{i.e.} open-closed boundary associated with an active region or quiet Sun, continuously open field deep inside a CH).  This level of specificity that moves beyond broad associations between the corona and solar wind (\textit{i.e.} coronal holes and faster wind, streamer belt and slower wind) is currently only achievable with coronal modeling.  In this work, we use the ADAPT-WSA model to identify the source locations of the observed solar wind for the first two \textit{Parker Solar Probe} encounters. We use ADAPT because this model flux evolves the photospheric magnetic field to account for areas of the Sun's surface that we do not observe (\textit{i.e.} polar regions and the solar far-side), and thus generates synchronic photospheric field maps (\textit{i.e.} for one moment in time) as opposed to traditional diachronic Carrington maps.  These photospheric field maps are then used as input to WSA to derive synchronic global coronal field solutions. The capabilities of ADAPT-WSA were particularly useful for the first two \textit{PSP} encounters because both occurred primarily on the Sun's far-side. Our modeling results are derived retrospectively in order to use photospheric field data from the encounter period (as opposed to forecasting the solar wind connectivity in advance), and validated with IMF and solar wind speed observations from \textit{PSP}. Our approach and tools (further discussed in Section~\ref{sec:2_Model}) allow us to provide the most realistic model solutions for each solar encounter.

There are several ways in which coronal modeling enabled the scientific interpretation and understanding of these encounters that would not have been possible with observations alone. Some of these points are highlighted below, and discussed in the context of our primary results.

\begin{itemize}
\item \textbf{Since the 3D coronal magnetic field is not easily nor routinely measured, the most reliable method for estimating the global coronal magnetic field configuration is coronal model extrapolations.}   For example, we were able to determine how the coronal magnetic field evolved in time while \textit{PSP} co-rotated with a far-side mid-latitude coronal hole in the first encounter (Figures~\ref{Fig_5:movie} \&~\ref{Fig_6:midlatCHevo}, discussed in \ref{sec3.2:enc1_discussion}). Since \textit{PSP} was on the solar far-side while co-rotating with the Sun, modeling this evolution is only possible using a model like ADAPT which uses flux transport processes evolves the photospheric magnetic field when observations are not available.  Our modeling allowed us to determine that the increase in solar wind speed and proton temperature from 2\,--\,9 November, 2018 (Fig.~\ref{Fig_3:PSP_data_with_WSA_1}) was likely because this mid-latitude coronal hole (\textit{i.e.} the model-derived solar wind source region) increased in total open area 1\,--\,8 November, allowing for faster, hotter, and less dense solar wind to form. While we are unable to validate the coronal hole areas derived by ADAPT-WSA with far-side corona observations, the model-derived $B_r$ and solar wind speed agree well overall for this encounter which gives us confidence in our results. 

\item \textbf{Coronal and solar wind models are currently the most accurate way to estimate the sources of the solar wind observed by spacecraft.} With ADAPT-WSA, we derive the source regions of the in situ observed solar wind down to the field line resolution of our model (\textit{i.e.} 2$^\circ$ in this study). We also quantify our results with empirical parameters (\textit{e.g} coronal hole boundary distance, expansion factor), and model-derived calculations such as the spacecraft separation from the HCS. The latter is a new WSA data product developed for \textit{PSP} encounters, yet it can be derived for  any spacecraft.  Quantifying the location of \textit{PSP} (or any in situ spacecraft) relative to the HCS helps us to identify when the spacecraft observes solar wind near to or far from the HCS, a necessary constraint for solar wind formation theories \citep{Antiochos2011}.   Our results show that most of the observed solar wind from both encounters originated from the boundaries of coronal holes when the spacecraft was near the HCS.  These periods have source regions with quiet Sun magnetic field (\textit{i.e.} no active regions present), higher expansion factors, and low coronal hole boundary distance. The observed solar wind for these periods is slow (300\,--\,400 km s$^{-1}$ on average). There is a short period in the first encounter (15\,--\,20 November, 2018) when \textit{PSP} observes moderately fast wind, far from the HCS (up to 20 degrees away as shown in Fig.~\ref{Fig_3:PSP_data_with_WSA_1}b), from deep inside a mid-latitude coronal hole (see Fig.~\ref{Fig_2:WSA_Enc1}a between $\sim$270\,--\,300$^\circ$ CL). Through this work, we demonstrate that we can more reliably determine when the in situ observed solar wind originates near to or far from the HCS, and the magnetic open-closed boundary. This capability can be used to investigate how populations of solar wind from these different sources are heated and accelerated. 

\item \textbf{Corona and solar wind models are necessary to distinguish between in situ variability caused by connectivity changes (\textit{i.e.} spatial structure in the corona) vs. time variability in the source region itself.}  For example, in the first encounter \textit{PSP} observed the solar wind from a mid-latitude coronal hole over an approximate 11-day period.  However, the ADAPT-WSA model output showed that for a subset of the co-rotation period (\textit{i.e.} 4\,--\,10 November, 2018), \textit{PSP} observed the solar wind from the same source region (\textit{i.e.} within a few model grid cells at 2$^\circ$ resolution, or roughly a few supergranules) within this mid-latitude coronal hole. Having this information allows one to interpret observations as being mostly time-dependent since the location of the source region is constant for several days. 

\end{itemize}

The results of this work highlight the current capabilities and limitations of both coronal modeling, and our ability to leverage multiple observatories to study the corona and solar wind.  For both encounters \textit{PSP} was located primarily on the solar far-side, where we currently do not have photospheric field observations to drive models.   While ADAPT flux evolves the photospheric field to mitigate the effects of not having far-side observations, ADAPT can only routinely incorporate new far-side flux emergence once it has been observed on the near-side. There are ways to include far-side emergence in ADAPT maps with helioseismology \citep{Arge2013}, however this is only possible on a case-by-case basis. \textit{Solar Orbiter}'s PHI instrument will provide the first photospheric field measurements of the solar far-side (albeit periodically), allowing us to drive coronal models with full-sun photospheric field observations for the first time. 

In the first encounter, we were able to obtain excellent agreement between the model-derived IMF and solar wind speed, and that observed at \textit{PSP} because there was not a significant flux emergence on the far-side during this period.  However, during the second encounter an active region formed on the solar far-side that was unknown to the model until it rotated onto the near-side on 9 April, 2019. This resulted in the model producing unreliable forecasts of the observed solar wind source regions for about 14 days.  Accurate forecasts are critical for when \textit{PSP} has a near-side solar encounter in order to coordinate simultaneous measurements of the solar wind and SEPs observed at \textit{PSP} with remote corona observations of their predicted source region. Nevertheless, the ADAPT-WSA solution stabilized and agreed well with \textit{PSP} observations about 2 days after the AR rotated onto the near-side and was incorporated into the model.  This indicates that models are able to produce accurate representations of the coronal field when all major flux emergence is accounted for at the time, demonstrating how critical it is for models to have global photospheric field observations. In this case, it was only possible to produce reliable model results for \textit{PSP}'s second solar encounter via retrospective analyses, as this active region would have been unknown to the model at the time of a forecast. Lastly, although the addition of the SCS model in WSA has been shown at times to produce better agreement with observations \citep{McGregor2011}, for these encounters, both WSA and standard PFSS models produced similar results \citep{Szabo2020,Badman2020,Panasenco2020}. This is likely due to these periods being relatively quiet. For these encounters, the use of a flux transport model like ADAPT paired with any coronal model made the most difference, because they are able to estimate the photospheric evolution on the far-side.

Furthermore, remote and in situ observatories were advantageously aligned during the first encounter to observe a streamer-blowout CME (discussed in~\ref{sec3.2:enc1_discussion}, and published in \citealt{Korreck2020,Nieves-Chinchilla2020}) that occurred on the Sun's far-side. Currently, it is only possible to simultaneously observe a transient event in both the corona and the solar wind by coincidence, as there is not a coordinated space-based mission in which one spacecraft observes the corona remotely and another observes the solar wind in situ, while both remain in quadrature.  Even with such a mission, the eruption would have to be orientated in such a way for the coronagraph to observe the event off the limb, and in the direction of the in situ spacecraft. Thus, both the accuracy of our modeling and the ability to fully characterize the SBO-CME in the first encounter were to some degree dependent on fortuitous circumstances (\textit{i.e.} no far-side AR emergence, random alignment of spacecraft).  Fortunately, models enable coordinated science between multiple missions by providing forecasts of where in situ spacecraft will observe the solar wind from, allowing ground-based observatories to point their instruments in these locations in advance. Leveraging models to coordinate observatories from multiple vantage points is increasingly imperative as new, advanced space and ground and space-based instruments come online to observe the Sun and solar wind in unexplored locations (\textit{i.e. Solo} at the poles and far-side) and at the finest spatial scales (\textit{i.e.} \textit{DKIST}).

 
\section{Conclusion} \label{sec:6_Summary}

In this work, we use the ADAPT-WSA model to retrospectively derive the global coronal magnetic field and solar wind source regions for the first two \textit{Parker Solar Probe} solar encounters.  Knowledge of where the observed solar wind came from back at the Sun enables deeper scientific understanding into the wind's formation (\textit{i.e.} how it was released, heated, accelerated), and allows one to interpret in situ variability either as connectivity changes (i.e. spatial structure in the corona) or time evolution of the solar wind source region itself. We present several new results for Encounters 1 and 2, including the time evolution of the far-side mid-latitude coronal hole that \textit{PSP} co-rotates with (Figures~\ref{Fig_5:movie} \&~\ref{Fig_6:midlatCHevo}).  As \textit{PSP} co-rotated with this coronal hole, the open area and model-derived solar wind speed increases, providing a possible explanation for the increase in speed and temperature observed at \textit{PSP} (Figure~\ref{Fig_3:PSP_data_with_WSA_1}).  We also explain in detail how new far-side AR emergence made Encounter 2 difficult to model, yet source regions of the solar wind can be reliably estimated once this AR rotates onto the near-side. The results discussed in this paper were vital to several science investigations that worked toward fulfilling the \textit{PSP} science objectives.  For example, they have been validated through comparison with several other PFSS and MHD models and shown to agree well overall with these solutions \citep{Szabo2020}. Further, model output from the first encounter was used to identify the source region of a CME observed at \textit{PSP}, characterize the eruption as a streamer blowout, all of which provided context to the analysis of the magnetic structure and plasma dynamics observed during this eruption \citep{Korreck2020,Nieves-Chinchilla2020}.  Our results show the importance of coordinated efforts between spacecraft at multiple vantage points, and the critical need for global (\textit{i.e.} including the solar polar regions and far-side), continuous observations of the solar atmosphere.  Photospheric observations of the Sun's polar regions and far-side are necessary to form complete inner boundary condition for models, and 4$\pi$ corona observations are necessary to validate and constrain model-derived open flux.


\acknowledgments

SW was supported by the NASA Postdoctoral Program; CNA and NMV were supported by the competed Heliophysics Internal Scientist Funding Model; 

This work utilizes data produced collaboratively between Air Force Research Laboratory (AFRL) and the
National Solar Observatory (NSO). The ADAPT model
development is supported by AFRL, along with AFOSR
(Air Force Office of Scientific Research) tasks 18RVCOR126 and 22RVCOR012. The input data utilized by ADAPT is obtained by NSO/NISP (NSO Integrated Synoptic Program). NSO is operated by the Association of Universities for Research in Astronomy (AURA), Inc., under a cooperative agreement with the National Science Foundation (NSF);

Parker Solar Probe was designed, built, and is now operated by the Johns Hopkins Applied Physics Laboratory as part of NASA’s Living with a Star (LWS) program (contract NNN06AA01C). Support from the LWS management and technical team has played a critical role in the success of the Parker Solar Probe mission;

Thank you to the FIELDS team for providing data (PI: Stuart D. Bale, UC Berkeley);

Thank you to the Solar Wind Electrons, Alphas, and Protons (SWEAP) team for providing data (PI: Justin Kasper, BWX Technologies).

\bibliography{ref}{}
\bibliographystyle{aasjournal}

\end{document}